\pdfoutput=1
\documentclass{article}
\usepackage{iclr2027_conference,times}
\usepackage[T1]{fontenc}
\usepackage[utf8]{inputenc}
\usepackage{amsmath,amssymb}
\usepackage{graphicx}
\usepackage{placeins}
\usepackage{wrapfig}
\usepackage{booktabs}
\usepackage{xcolor}
\usepackage{hyperref}
\usepackage{url}

\graphicspath{{figures/}}
\hypersetup{
  pdftitle={Music Circuit: Trace and Control Style Dynamics in Large Music Generation Model},
  pdfauthor={Ruiji Yu, Wenye Ma, Gus Xia}
}

\title{Music Circuit: Trace and Control Style Dynamics in Large Music Generation Model}
\author{Ruiji Yu$^{1}$, Wenye Ma$^{2}$ \& Gus Xia$^{1}$ \\
$^{1}$Mohamed bin Zayed University of Artificial Intelligence \quad
$^{2}$McGill University \\
\texttt{\{ruiji.yu,gus.xia\}@mbzuai.ac.ae}, \texttt{wenye.ma@mail.mcgill.ca}}

\iclrfinalcopy
\usepackage{etoolbox}
\makeatletter
\patchcmd{\@maketitle}{Published as a conference paper at ICLR 2027}{Preprint}{}{\errmessage{Preprint header patch failed}}
\makeatother

\begin{document}
\maketitle

\begin{abstract}
Musical style, a global attribute that functions throughout a whole piece, is a lever widely used by users in modern large music generation models. Yet how
style is represented inside these models, and which components drive it,
remains largely unexplored. In language models, circuit analysis traces a
representation to a sparse sub-network of a few key components, namely attention
heads and MLP neurons, and has been used successfully to explain and control
global attributes such as emotion. In this work, we use circuit
analysis to study the mechanisms behind musical style. Based on YuE2, we
build a dataset that pairs the same lyrics with five styles and design a
three-step framework. First, stage localization shows that the style forms
mainly in the score-planning stage. Second, static observation across the positions of
this stage further reveals that style is not a fixed vector but a dynamic
representation across tokens. Third, we therefore merge the activation differences across positions
into one direction per style, and trace the neurons and attention heads that
drive this direction into a sparse circuit. Without any style text, steering this circuit successfully transfers one piece into the target style, matching and even exceeding the style prompt. To
our knowledge, this is the first work to analyze and control
musical style at the circuit level in a large music generation model.
Audio and score examples are available at \url{https://aristo233.github.io/Music_Circuit/}.

\end{abstract}

\section{Introduction}
\label{sec:introduction}

\suppressfloats[t]
\begin{figure}[t]
\centering
\includegraphics[width=\linewidth]{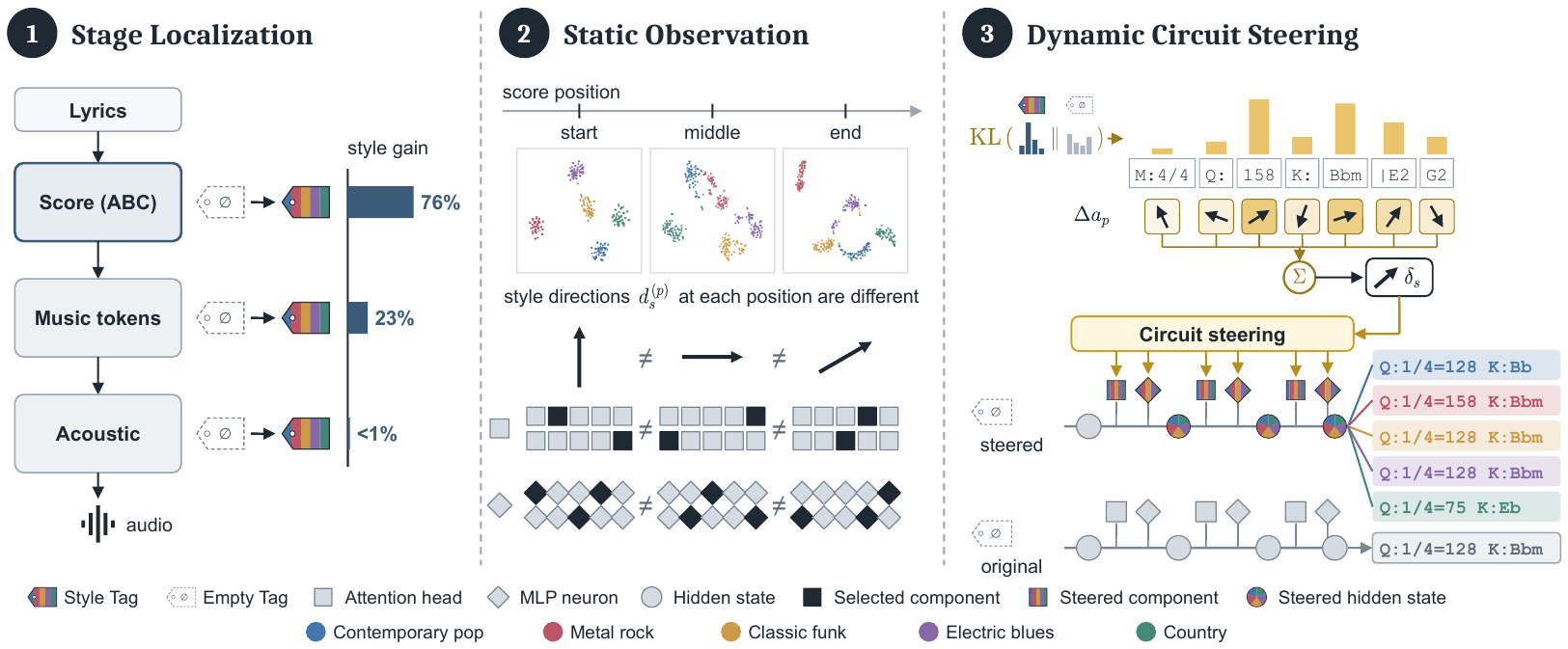}
\caption{Overview. (1) \textbf{Stage localization}: swapping the empty tag for
the style tag in one generation stage at a time shows that most of the style
gain comes from the score stage. (2) \textbf{Static observation}: styles are separable at every score position,
but the style direction $d_s^{(p)}$ (layer index omitted) and the attention heads
and MLP neurons that contribute most to it change from position to position.
(3) \textbf{Dynamic circuit steering}: the activation differences $\Delta a_p$
between the two tags are weighted by the per-step KL divergence between their
next-token distributions and summed into one direction $\delta_s$ per style,
which steers the circuit. Without any style text, the circuit writes the same
lyrics in five styles, matching and often exceeding the style prompt; the ABC
headers on the right are real generations for one test lyric.}
\label{fig:overview}
\end{figure}

Music generation models are advancing rapidly.\footnote{Code:
\url{https://github.com/Aristo233/Music_circuit_code}.} Symbolic models write note events
or text scores such as ABC notation token by
token~\citep{huang2019musictransformer,yu2022museformer,yuan2024chatmusician,qu2024mupt,wang2025notagen},
while audio models generate discrete acoustic tokens and use text to control
attributes such as genre and
instrumentation~\citep{dhariwal2020jukebox,agostinelli2023musiclm,copet2023musicgen}.
In controllable music generation, style is one of the
most important attributes that users control. Music style transfer~\citep{dai2018styletransfer} keeps the content, such as the lyrics or
the melody, and changes the overall character of a piece in its mode, rhythm,
instrumentation, timbre and vocal delivery. Genre is one of the main perspectives
on musical style. Yet this control still relies mainly on text prompts, and we
know little about how style, as a global representation, works inside the model.

Mechanistic interpretability research offers tools for such questions: by analyzing a
model's internal representations, one can find directions that correspond to
specific concepts and steer generation directly along
them~\citep{bereska2024mechanistic,turner2023actadd,zou2023repe,rimsky2024caa}.
In music models, prior work shows that style can likewise be extracted as a
direction in activation space and used for steering, for example as composer
vectors~\citep{jiang2026composervector} or as musical concepts found by sparse
autoencoders~\citep{singh2026concepts}.

Although effective in practice, these methods operate only on directions in the residual stream. They
do not explain how such a style direction is formed by specific internal components. Circuit methods offer a way to answer these questions: they
decompose the behavior-specific computation into an internal sparse sub-network with important components, namely
attention heads and MLPs~\citep{elhage2021framework,wang2023ioi,conmy2023acdc}. 
Such methods have been used to explain and control global attributes similar to
style. Studies of emotion find readable emotion representations inside the model
and show that intervening on them changes the model's
behavior~\citep{wang2025emotioncircuits,anthropic2026emotions}. By intervening on the
components that substantially drive the emotion representation,
\citet{wang2025emotioncircuits} produce replies with clearly different emotions
for the same scenario without any emotional prompt. This can be seen as a form
of style transfer in natural language. Circuit analysis, however, has seen very
limited use in the study of musical style.

We therefore ask how style is represented in large music generation models, and decompose
the question into three parts. (1) In what form does style, a dynamic global concept, exist in the model's representations? 
(2) Which internal mechanisms and components substantially drive this representation? (3) Can
these components be used to control style effectively?

To answer these questions, we propose a style circuit discovery and control
approach based on YuE2~\citep{yuan2025yue}, a state-of-the-art open-source music generation model. It first plans a melody as a score, then generates music tokens and
renders them as audio. We first build a dataset in which the same lyrics are generated
in five styles, namely contemporary pop, metal rock, classic funk, electric blues
and country (\S\ref{sec:setup}). On this basis, we design a three-step framework
(Figure~\ref{fig:overview}; \S\ref{sec:methodology}).
(1) \textbf{Stage localization.} Ablating the style tag separately in each of the
three generation stages shows that most of the style effect is formed during
score planning.
(2) \textbf{Static observation.} We extract style directions and select
components at multiple positions of the score and the music tokens, and check whether they stay stable
during generation. We find that style is clearly separable at every position, but the
directions and the corresponding important components keep changing with position and barely
overlap. We therefore hypothesize that global style is determined not by a static vector at a fixed
position, but by a dynamic representation across tokens.
(3) \textbf{Dynamic circuit steering.} Accordingly, we weight each step by the KL
divergence between the output distributions under the style tag and under the
empty tag, and aggregate the activation differences across positions into one
direction that captures the style concept that stays invariant during dynamic
generation. We then trace the neurons and attention heads that drive this
direction, assemble them into a sparse circuit, and test their role through
ablation, enhancement, and steering generation toward a target style.
Without any style text, this circuit intervention matches and often exceeds the
style prompt: under CLAP it hits the target style in 133 of 150 generations on
the 15 test lyrics against 113 for the prompt, while a random circuit with the
same number of components and injection norm stays at the empty-tag baseline
(34; \S\ref{sec:exp-steering}).

Our contributions are threefold.
(1) \textbf{Insightful understanding of musical style representation.} Taking genre as the entry point to
musical style, analogous to emotion as a concept in natural
language, we find that style in music generation models is stored dynamically:
it can be read out at every position, but is carried by directions and
components that change with position. This fits the way music organizes
information at several levels, such as mode, rhythm and instrumentation.
(2) \textbf{Circuit-level control.} We propose a KL-weighted dynamic style
direction and its sparse circuit, which together control music generation at the circuit
level without any style text.
(3) \textbf{Mechanistic evidence.} We take a first step toward component-level mechanistic interpretability
of musical style.

\section{Related Work}
\label{sec:related-work}

\subsection{Music Generation Models}
\label{sec:related-music}

Symbolic models generate note events or text-based scores such as ABC
notation~\citep{huang2019musictransformer,yu2022museformer,lu2023musecoco,yuan2024chatmusician,qu2024mupt,wang2025notagen},
while audio models generate acoustic tokens or latents and use text to control
attributes such as genre and
instrumentation~\citep{dhariwal2020jukebox,huang2023noise2music,agostinelli2023musiclm,copet2023musicgen,liu2024audioldm2,evans2024stableaudio,yuan2025yue}.
Style has been transferred or controlled with dedicated models at both
levels~\citep{brunner2018midivae,cifka2020groove2groove,wu2024musiccontrolnet},
and genre is its standard label in music information
retrieval~\citep{tzanetakis2002genre}. Prior interpretability work has probed the internal representations of music models
\citep{castellon2021jukemir,ma2024probing},
intervened on internal activations for steering
\citep{ma2024probing,facchiano2025patching},
and used sparse autoencoders to discover interpretable musical concepts
\citep{singh2026concepts},
but these approaches do not trace the specific components that write this information.

\subsection{Circuit Discovery and Mechanistic Interpretability}
\label{sec:related-circuits}

Mechanistic interpretability aims to reverse-engineer model computations
into human-understandable mechanisms~\citep{bereska2024mechanistic}.
Circuit discovery, a central approach, decomposes Transformers into attention
heads and MLPs that read from and write to the residual
stream~\citep{elhage2021framework,geva2021ffn,olsson2022induction} and locates
circuits for specific behaviors with causal
interventions~\citep{vig2020causal,geiger2021causal,meng2022rome,wang2023ioi,hanna2023greaterthan}
and automated methods~\citep{conmy2023acdc,syed2024attribution,ferrando2024ifr}.
Another line of work finds linear concept
directions~\citep{nanda2023othello,park2024linear} or sparse
features~\citep{bricken2023monosemanticity,cunningham2024sae} and steers
generation by adding them to the residual
stream~\citep{turner2023actadd,zou2023repe,rimsky2024caa}. Emotion
Circuits~\citep{wang2025emotioncircuits} combine the two: they extract
emotion directions, then locate and intervene on the components that
write along them. Circuit discovery in music generation models remains largely unexplored.

\subsection{YuE2}
\label{sec:preliminary}

YuE2~\citep{yuan2025yue}\footnote{YuE2 architecture and inference details:
\url{https://huggingface.co/m-a-p/YuE2-3B}.}
is a conditional music generation model that combines symbolic score
planning with audio generation. Lyrics provide the words to sing,
a style text describes genre, instrumentation, and vocal characteristics,
and optional scores specify melody, rhythm, and chords. YuE2 uses a
Mixture-of-Transformers architecture with autoregressive (AR) and
non-autoregressive (NAR) branches. When score planning is enabled, the AR
branch first generates an ABC score. It then predicts music tokens
based on the text and score. The NAR branch uses these tokens and the
conditioning context to generate acoustic latents through flow matching.
A variational autoencoder (VAE) decoder then converts these latents into
audio. YuE2 also supports user-provided scores and generation without
score planning. Its style tags can affect all three stages, which we call the
score-planning (score) stage, the music-token stage and acoustic rendering. This is consistent with the broad meaning
of musical style in practice, which includes melodic motifs, harmonic color,
instrumentation, playing techniques, and vocal expression.

\section{Methodology}
\label{sec:methodology}

Our method has three steps (Figure~\ref{fig:overview}): stage localization,
static observation and dynamic circuit steering. First, a stage-wise tag ablation
(\S\ref{sec:stage-ablation}) shows that about three quarters of the style is
formed in the score stage (\S\ref{sec:exp-stage}), so the subsequent analysis
and interventions focus on this stage. Second, a single-position analysis (\S\ref{sec:single-position}) shows that
style can be read out at every position, but the directions and the components
that push along them differ across score positions (\S\ref{sec:exp-distribution}), and
the style tag changes the output substantially at only a few positions
(Appendix~\ref{app:cross-position}); no single position represents the whole
generation. Third, we aggregate activation differences across positions into a
dynamic direction, weighting each step by its KL divergence, and build a sparse
circuit from it (\S\ref{sec:klw-steering}). These components act causally
(\S\ref{sec:exp-causal}) and, without any style text, match or exceed the style
prompt (\S\ref{sec:exp-steering}).

\subsection{Problem Setup and Notation}
\label{sec:setup-notation}

We work with songs that YuE2 generates for the same lyrics in five styles
(\S\ref{sec:setup}). Directions, components and intervention settings are
estimated only on the representation analysis set $\mathcal I_{\mathrm{ext}}$ (the set from which directions are extracted);
the test set is used only for evaluation.

YuE2 generates a song in the score stage, the music-token stage and acoustic
rendering (\S\ref{sec:preliminary}), and each stage can read the style tag. We write $c_s$ for the tag
prefix of style $s$ and $c_\varnothing$ for the empty tag prefix.
To extract representations, we feed a song's prefix (style tag and lyrics),
together with the score and music tokens it has already produced, into the
autoregressive model in one forward pass without sampling, and read the
internal states at every layer and position. When we compare the two tags, both
forward passes share the same score history $x$ and only the prefix changes, so
any difference comes from the style tag alone. At layer $\ell$ and position $p$,
$h_p^{(\ell)}$ denotes the residual stream and $a_p^{(\ell)}$ the component
activations: for MLP neurons, the gated activation entering the down projection
$W_{\mathrm{down}}^{(\ell)}$; for attention heads, the head output entering the
output projection $W_O^{(\ell)}$.

\subsection{Stage Localization: Stage-wise Tag Ablation}
\label{sec:stage-ablation}

The style tag of each of the three stages is set independently to $c_s$ or
$c_\varnothing$; when a later stage's tag is switched, the outputs of the earlier
stages are reused, so each stage's effect on the final audio can be isolated.
We measure audio style with the CLAP~\citep{wu2023clap} margin
(\S\ref{sec:setup}). Because the gain from one stage's tag depends on whether
the other stages carry it (once the score carries the style, the music-token tag
helps less), we split the margin gain of the full tag over the empty tag among
the stages by their Shapley values~\citep{shapley1953value}, which sum exactly
to the total gain (Appendix~\ref{app:stage-ablation}).

\subsection{Static Observation: Single-position Style Directions and Circuits}
\label{sec:single-position}

\paragraph{Style direction.} To separate style from the rest of the song's
content, we take the five style versions of the same lyrics and subtract their
mean hidden state from each sample, which removes what the versions share,
chiefly the lyrics, and keeps what differs by style. Summing over lyrics and
normalizing gives the direction of style $s$ at layer $\ell$ and position $p$:
\begin{equation}
d_s^{(\ell,p)}\propto\sum_{i\in\mathcal I_{\mathrm{ext}}}\Big(h_{i,s,p}^{(\ell)}-\frac{1}{5}\sum_{s'=1}^{5}h_{i,s',p}^{(\ell)}\Big),
\qquad \big\|d_s^{(\ell,p)}\big\|_2=1 .
\label{eq:single-direction}
\end{equation}

\paragraph{Component attribution.} Each MLP neuron writes to the residual stream
through one column of the down-projection matrix, and each attention head writes
its output through its column block of the output projection. A component's
contribution to the style direction depends on both its activation and its write
direction:
\begin{equation}
c_{s,j}^{(\ell,p)}=g_{s,j}^{(\ell,p)}\,\big\langle W_{\mathrm{down},:,j}^{(\ell)},\,d_s^{(\ell,p)}\big\rangle ,
\qquad
c_{s,h}^{(\ell,p)}=\big\langle W_{O,h}^{(\ell)}\,z_{s,h}^{(\ell,p)},\,d_s^{(\ell,p)}\big\rangle ,
\end{equation}
where $g_{s,j}^{(\ell,p)}$ is the mean (uncentered) gated activation of neuron
$j$ over the songs of style $s$, $z_{s,h}^{(\ell,p)}$ is the mean output of head
$h$, and $W_{O,h}^{(\ell)}$ is the column block of the output projection for that
head. $d_s^{(\ell,p)}$ is fitted at the sublayer output where the component
writes: after attention for heads, after the MLP for neurons. In each layer we select
components in descending order of contribution.

\paragraph{Causal validation.} We test the selected components $J$ with ablation
and enhancement. Centering the component activations in the same way and
averaging over the representation analysis set gives a style difference
$\bar\delta_s^{(\ell)}$ in component space. Ablation zeroes the selected
components under the style tag; enhancement adds this difference to them under
the empty tag:
\begin{equation}
\text{ablation: } a_J^{(\ell)}\leftarrow 0,
\qquad
\text{enhancement: } a_J^{(\ell)}\leftarrow a_J^{(\ell)}+\bar\delta_{s,J}^{(\ell)} .
\end{equation}
We measure the effect as the summed change of the residual projections on the
style direction across layers, reading the residual after each sublayer and
projecting it on the direction fitted at that sublayer output,
\begin{equation}
\Delta S=\sum_{\ell=1}^{28}\big\langle \widetilde h^{(\ell)}-h^{(\ell)},\,d_s^{(\ell)}\big\rangle ,
\end{equation}
where $\widetilde h$ is the residual after the intervention (the position index
is omitted), and compare with random components of the
same number per layer and, for enhancement, the same injection norm.

\paragraph{Cross-position comparison.} We extract directions and select
components separately at several score positions, and compare positions by the
cosine between directions and the Jaccard overlap of selected components. If
style were a static vector, they would agree across positions. The circuit built
from one position also serves as a baseline in the generation experiments.

\subsection{Dynamic Circuit Steering with KL Weights}
\label{sec:klw-steering}

\paragraph{Position weights.} Since style representations change with position,
we need each position's share of the style difference over the whole piece. The
style tag does not change the output equally at every position, so we measure
its effect at each position by the stepwise KL divergence
\begin{equation}
\mathrm{KL}_p=\mathrm{KL}\big(p(\cdot\mid c_s,x_{<p})\,\|\,p(\cdot\mid c_\varnothing,x_{<p})\big).
\end{equation}
By the chain rule, the KL divergence between the two conditions' distributions
over the whole generation equals the expected sum of the stepwise conditional KL
divergences:
\begin{equation}
\mathrm{KL}\big(p(x_{1:T}\mid c_s)\,\|\,p(x_{1:T}\mid c_\varnothing)\big)
=\mathbb E_{x\sim p(\cdot\mid c_s)}\sum_{t=1}^{T}\mathrm{KL}_t .
\end{equation}
KL therefore gives each step's share of the overall style difference: KL tells
where style acts, the activation difference tells in which direction.

\paragraph{KL-weighted direction.} Within each lyric group, we average the
component activation differences between the style tag and the empty tag,
$\Delta a_{i,s,p}^{(\ell)}=a_{i,s,p}^{(\ell)}-a_{i,\varnothing,p}^{(\ell)}$, with
KL weights, then average over lyrics:
\begin{equation}
\delta_s^{(\ell)}=\frac{1}{|\mathcal I_{\mathrm{ext}}|}\sum_{i\in\mathcal I_{\mathrm{ext}}}
\frac{\sum_p \mathrm{KL}_{i,s,p}\,\Delta a_{i,s,p}^{(\ell)}}{\sum_p \mathrm{KL}_{i,s,p}}.
\label{eq:klw}
\end{equation}
Applying the same weighted average to the residual stream after each sublayer
and normalizing gives a unit direction $u_s^{(\ell)}$ in residual space. Each style thus has one direction that stays fixed throughout the score stage:
the KL-weighted aggregate of the position-specific differences, i.e., the part of
the style representation shared by the positions that matter
(Appendix~\ref{app:klw-details}).

\paragraph{Circuit construction.} The single-position score measures how much a
component accounts for in the style representation. For steering we care instead
about where the injected quantity pushes the residual, so we score each
component by its KL-weighted difference $\delta_s$, which is exactly what will
be injected: a neuron's score is its difference component times the projection
of its write direction on $u_s^{(\ell)}$, and a head's score is the projection of
its difference, after the output projection, on $u_s^{(\ell)}$, each taken at the
sublayer output where the component writes
(Appendix~\ref{app:klw-details}). A positive score means that injecting the
component's difference pushes the residual along the style direction in its own
layer, and we keep only such components.

\paragraph{Causal validation.} We reuse the single-position ablation and
enhancement, but both interventions now act on every position of the score:
ablation zeroes the selected components under the style tag, and enhancement
adds $\delta_s$ to them under the empty tag. The change of the residual
projection on $u_s$ is averaged over positions with each lyric's own stepwise KL
weights (Appendix~\ref{app:klw-details}). These weights are used only for
evaluation; generation always injects the same fixed direction for each style.
Random controls are drawn only from components that the target does not select
(Appendix~\ref{app:random-controls}).

\paragraph{Circuit assembly.} Each style's circuit has 560 components per stage
(ten per sublayer), 392 MLP neurons and 168 attention heads. In the score stage
we rank all layers together, since the KL-weighted scores already reflect how
much each layer writes; in the music-token stage the budget is spread across
layers by sublayer importance~\citep{wang2025emotioncircuits}
(Appendix~\ref{app:sublayer-importance}). Components are selected once per style
and shared by all lyrics.

\paragraph{Steering.} During generation, at every step of the score stage we add
the KL-weighted difference, scaled by a strength $\alpha$, to the selected
components $J$ and leave everything else unchanged:
\begin{equation}
a_J\leftarrow a_J+\alpha\,\delta_{s,J}.
\end{equation}
Because the style direction changes with position and style decisions are spread
over the whole score, we inject the same direction throughout the stage rather
than only at its start. The music-token stage injects, in the same way, the difference averaged with
equal weights over its first 128 steps; all circuit conditions and CAA share this
music-token intervention. Acoustic rendering is not intervened on, and no
stage receives any style text.

\section{Experiments}
\label{sec:experiments}

\subsection{Setup}
\label{sec:setup}

\begin{wrapfigure}{r}{0.48\linewidth}
\vspace{-12pt}
\centering
\includegraphics[width=\linewidth]{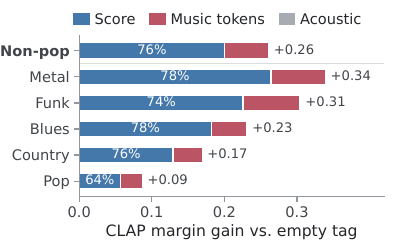}
\caption{Stage-wise style-tag ablation (54 representation-analysis lyrics):
Shapley decomposition of the full-tag CLAP effect; percentages give the score
stage's share.}
\label{fig:stage-factorial}
\vspace{-10pt}
\end{wrapfigure}
\paragraph{Dataset.} We pair short lyrics written with GPT-6 Astra~\citep{openai2026gpt6astra} with prompts
for five styles distinct in rhythm, instrumentation and vocals: contemporary pop,
metal rock, classic funk, electric blues and country (pop, metal, funk, blues and
country below). YuE2 generates the ABC scores and audio, and we keep only lyric
groups whose five versions all pass CLAP and Gemini 3.6
Flash~\citep{googledeepmind2026gemini36flash} checks. The 69 lyric groups are split
into a representation analysis set $\mathcal I_{\mathrm{ext}}$ (54 groups) and a test
set (15 groups; Appendix~\ref{app:data}).

\paragraph{Evaluation.} Two automatic judges assess style. CLAP scores the audio
against the five style descriptions; the \emph{margin} is the target style's
score minus the best other style's, and a complete generation with a positive
margin is a \emph{hit}. Gemini 3.6 Flash picks a style blind among the five styles and a distractor,
and a hit is picking the target. Truncated generations are misses. A subjective evaluation complements them.

\paragraph{Baselines.} We compare with the empty-tag baseline; the style prompt,
with style text in all three stages; a random circuit, which matches the
dynamic circuit in recipe, timing and per-layer injection norm but uses
random components disjoint from the target; and CAA~\citep{rimsky2024caa}, conventional steering that adds the mean residual
difference between the style tag and the empty tag at one layer at every score
step. CAA and the circuits share the dynamic circuit's music-token
intervention and differ only in the score stage. The strength $\alpha=0.5$ and
CAA's layer and strength are chosen on tuning lyrics of the representation
analysis set (Appendices~\ref{app:implementation}
and~\ref{app:audio}).

\subsection{Style forms mainly during score planning}
\label{sec:exp-stage}

The stage-wise ablation shows that the score stage accounts for about three
quarters of the full style-tag effect, the music-token stage for one quarter,
and acoustic rendering for almost none
(Figure~\ref{fig:stage-factorial}). Two further observations agree: the style tag
changes global score settings such as key, tempo and meter, and replacing only
the empty-tag score with the score written under the style tag moves the audio
clearly toward the target style (Appendix~\ref{app:data}). We therefore extract
directions and intervene in the score stage, with the music-token stage as a
complement.

\subsection{Style is encoded dynamically across score positions}
\label{sec:exp-distribution}

\begin{figure}[t]
\centering
\includegraphics[width=0.88\linewidth]{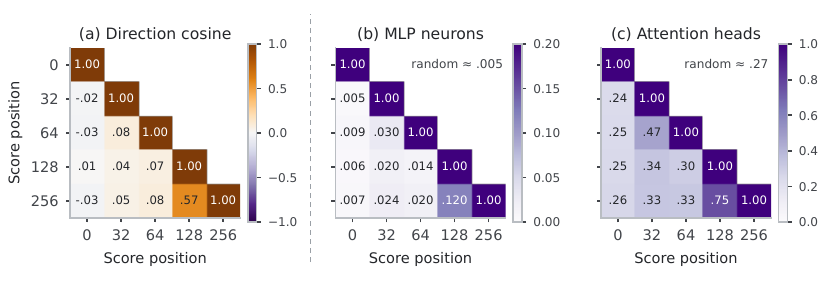}
\caption{Across score positions at layer 28 (mean of five styles): (a) cosine
between directions; (b, c) Jaccard overlap of the selected MLP neurons and
attention heads (chance 0.005 and 0.27).}
\label{fig:cross-position}
\end{figure}

After within-lyric centering, the five styles separate at the start and end of
the score and at the start of the music tokens; mid-score, where hidden states
barely distinguish styles, attention-head outputs and MLP activations still do
(Appendix~\ref{app:single-direction}). At each position, the direction is reproducible (Appendix~\ref{app:cross-position}),
and at the music-token positions its components are causal
(Appendix~\ref{app:single-circuit}). Score steps, however, barely share them (Figure~\ref{fig:cross-position}, at 0
to 256 generated tokens): the direction at the start of the score is nearly
orthogonal to later ones, its MLP neurons barely overlap with those of
other positions, and attention heads overlap only at chance level; only later
positions are clearly similar. Style is thus not a static vector fixed in one place, but a dynamic
representation that changes as the score is written (music-token directions are
more consistent after the first position, Appendix~\ref{app:cross-position}).

\subsection{A sparse circuit causally controls style}
\label{sec:exp-causal}

\begin{figure}[t]
\centering
\includegraphics[width=0.88\linewidth]{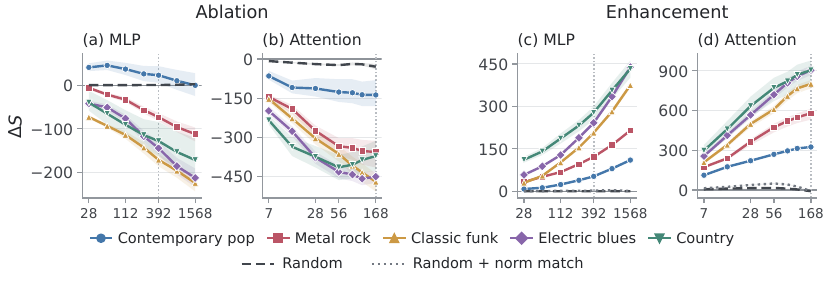}
\caption{Ablation and enhancement of MLP neurons and attention heads against the number
of edited components; vertical line: generation budget; bands: 95\% intervals.
Dashed gray: random components of the same number per layer; dotted gray: the
same with the target's injection norm.}
\label{fig:klw-causal}
\end{figure}

With the KL-weighted direction as reference, zeroing only 0.23\% of the MLP
neurons lowers the style projection and moves the output about 80\% as far from
the style-tag output as removing the tag, while injecting the difference pushes
the representation toward the target style without a tag
(Figure~\ref{fig:klw-causal}). Attention heads behave alike; a few injected heads recover most of the output
difference and more recover less, hence the small generation strength. Random components, even norm-matched, barely
matter, except that zeroing heads damages the output indiscriminately. Pop, the default style, has a weak direction whose projection rises under ablation
(Appendix~\ref{app:random-controls}).

\subsection{Circuit steering writes the target style without any style text}
\label{sec:exp-steering}

All conditions start from the lyrics and write their own score and music without
any style text; each of the 15 test lyrics and five styles is generated with two
seed pairs, 150 generations per condition.

\begin{table}[t]
\centering
\footnotesize
\setlength{\tabcolsep}{2.2pt}
\setlength{\abovecaptionskip}{0pt}\setlength{\belowcaptionskip}{5pt}
\caption{Hits per style on the test set under the two judges, out of 30 (15 lyrics,
two seed pairs); truncated generations are misses. Bold: best per column.}
\label{tab:steering}
\begin{tabular}{lrrrrrr@{\hspace{12pt}}rrrrrr}
\toprule
& \multicolumn{6}{c}{CLAP} & \multicolumn{6}{c}{Gemini 3.6 Flash} \\
\cmidrule(lr){2-7} \cmidrule(lr){8-13}
Condition & Pop & Metal & Funk & Blues & Country & Total & Pop & Metal & Funk & Blues & Country & Total \\
\midrule
Empty-tag baseline & 27 & 0 & 0 & 0 & 3 & 30 & 14 & 1 & 0 & 0 & 15 & 30 \\
Style prompt & 29 & 17 & 13 & \textbf{29} & 25 & 113 & 24 & 12 & 15 & \textbf{30} & \textbf{30} & 111 \\
Random circuit & 27 & 0 & 0 & 3 & 4 & 34 & 13 & 0 & 0 & 0 & 20 & 33 \\
CAA & 26 & 22 & \textbf{26} & 21 & 23 & 118 & 24 & 22 & \textbf{25} & 23 & 26 & 120 \\
Dynamic circuit (ours) & \textbf{30} & \textbf{24} & 25 & 27 & \textbf{27} & \textbf{133} & \textbf{25} & \textbf{23} & 22 & 27 & 29 & \textbf{126} \\
\bottomrule
\end{tabular}
\end{table}

Table~\ref{tab:steering} gives the hits per style under both judges. The two
judges agree: without any style text, the dynamic circuit hits more often than
the style prompt. The gap lies in metal and funk, where the prompt hits only about
half of the generations, its misses mostly heard as pop or blues, and the circuit about 80\%. On blues and country the prompt is near ceiling, and the circuit is slightly
behind on blues. The random circuit performs like the empty-tag baseline, so the effect comes
from the selected components, not the injection itself; circuits from
single-position or equally averaged directions hit far less often
(Appendix~\ref{app:steering-extra}), so the KL-weighted aggregation is key.

Compared with CAA, the circuit's advantage lies mainly in stability. Adding the
same fixed vector at every score step often derails generation: CAA truncates 23 of 150 generations against 4 for the circuit, and runs twice as
long (177 against 84 seconds); among complete generations its hit rate is no lower than the circuit's, but with
truncations as misses its total is significantly lower under CLAP and slightly
lower under Gemini. The circuit is also longer than the prompt (66 seconds), but
the random circuit is not (58 seconds), so injection size is not the cause.

\begin{figure}[t]
\centering
\includegraphics[width=0.95\linewidth]{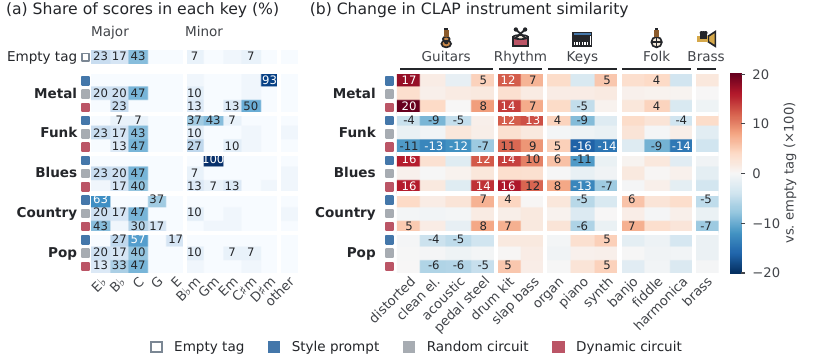}
\caption{Key and instrumentation on the test set. (a) Percentage of each row's
30 scores (15 lyrics, two seeds) in each key; the empty tag has one row. (b) CLAP
similarity to each instrument description minus that of the empty-tag audio of
the same lyric and seed, averaged and multiplied by 100.}
\label{fig:score-change}
\end{figure}

\paragraph{What the circuit changes, musically.} We ask which musical decisions
the circuit steers: the key, set in the score, and the instrumentation, set in
the music tokens (Figure~\ref{fig:score-change}). Each style has its typical
keys in the dataset (D$\sharp$m for metal, minor keys such as Gm and Dm for
funk and blues, major keys such as G and E$\flat$ for country, F and D for
pop), whereas the empty tag's key does not depend on the style
(Appendix~\ref{app:instrumentation}). The prompt puts nearly every score in the
typical key (93\% D$\sharp$m for metal, all Gm for blues). The circuit moves keys
in the same direction but less far: half of the metal scores land in
C$\sharp$m and country mostly in E$\flat$, while four in ten funk and blues
scores keep the empty tag's key, and the circuit never writes the 6/8 meter of
the blues prompt; the random circuit leaves keys unchanged.

Instrumentation, measured with CLAP instrument tags, is where the circuit acts
most like the prompt: distorted guitar, heavy drums and slap bass for metal and
blues, slap bass without acoustic guitar and piano for funk, pedal steel guitar
and banjo for country. Over 13 instruments, the changes of the circuit and the
prompt correlate at 0.92--0.98 across styles, against 0.46--0.76 for the
near-zero changes of the random circuit. The circuit thus realizes metal and
country through both the score and the instrumentation, and funk and blues
mainly through the instrumentation.

\begin{figure}[t]
\centering
\includegraphics[width=0.85\linewidth]{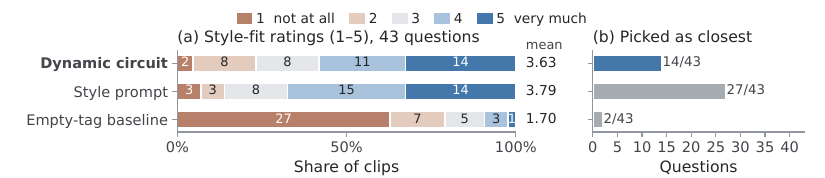}
\caption{Subjective evaluation (preliminary, 43 non-pop questions); bars count clips.}
\label{fig:listening}
\end{figure}

\paragraph{Subjective evaluation.} Listeners rate three anonymized clips of the same
lyric and seed (baseline, prompt and circuit) for style fit from 1 to 5 and pick
the closest (Figure~\ref{fig:listening}; Appendix~\ref{app:listening}).
Preliminary results (eleven listeners) agree with the automatic judges that the
circuit writes the target style: circuit and prompt are both rated well above the
baseline; between the two, listeners so far favor the prompt on the non-pop styles and
the circuit on pop, but neither difference is significant and both conditions
are rated far closer to each other than to the baseline.

\section{Conclusion}
\label{sec:conclusion}

We studied musical style in large music generation models at the circuit level.
With YuE2, we find that style forms mainly in the score-planning stage and is
stored dynamically across the tokens of the score: it can be read out at every
position, yet it is carried by directions and components that change with
position. Building on
this finding, we weight each step by its KL divergence to obtain one dynamic
direction per style, and trace the sparse circuit that drives it. These few
neurons and attention heads act causally in both ablation and enhancement, and
steering them directly matches or exceeds the style prompt in controlling
style, without any style text. The analysis already transfers to vocal gender, with an attribute-specific
aggregation (Appendix~\ref{app:vocal-gender}); we will extend it to further attributes such
as instrumentation and emotion and to other models, and trace how information
flows among the components of the circuit.

\bibliography{references}
\bibliographystyle{iclr2027_conference}

\clearpage
\appendix
\renewcommand{\topfraction}{0.9}
\renewcommand{\bottomfraction}{0.8}
\renewcommand{\textfraction}{0.07}
\renewcommand{\floatpagefraction}{0.8}
\setcounter{topnumber}{3}
\setcounter{bottomnumber}{2}
\setcounter{totalnumber}{5}
\section{Data Construction and Generation Protocols}
\label{app:data}

We construct a music dataset with matched lyrics across styles to study
style representations in YuE2. We first use GPT-6 Astra to generate 32 base
lyric texts, then create four narrative variants for each lyric family,
yielding 128 four-line English lyric texts. Each text is paired with
prompts for contemporary pop, metal rock, classic funk, electric blues,
and country. With score planning enabled, YuE2 first generates an ABC score from the lyrics
and the style text. It then generates music tokens and audio using the same
lyrics and style text together with the resulting score. Samples within each group share exactly the same
lyrics, while scores are generated independently for each style.
Generation continues until the audio ends naturally, and we retain
the complete 48\,kHz stereo output.

We then verify whether the generated samples match their target styles.
For CLAP, we combine scores from 10-second windows spanning each track,
weighted by their coverage. The target style must rank first among the
candidate descriptions, which are the five styles and a distractor, Chinese
traditional, with a similarity margin of at least 0.03 over
the highest-scoring alternative. For lyric--style combinations missing
after the initial filtering, we generate additional ABC scores and
complete audio tracks, retaining the first candidate that passes the
CLAP threshold. This gives 640 candidate samples. This filtering criterion is stricter than the
hit used to evaluate steering (\S\ref{sec:setup}), which compares only the five
style descriptions and requires a positive margin without a threshold.

We next use Gemini 3.6 Flash~\citep{googledeepmind2026gemini36flash} to
classify each of the 640 complete audio tracks, choosing among the five styles
and the same distractor.
We use anonymous
filenames and randomize the order of the style options. A total of
569 samples pass both the Gemini verification and the original CLAP
threshold. To support comparisons across styles using the same lyrics,
we retain only complete groups for which all five style versions pass
filtering. The final dataset contains 69 lyric groups, with 69 songs
per style and 345 songs in total.

\subsection{Data Construction Example}
\label{app:data-example}

\begin{figure}[htbp]
\centering
\begingroup
\definecolor{casebackground}{gray}{0.96}
\setlength{\fboxsep}{8pt}
\colorbox{casebackground}{%
\begin{minipage}{\dimexpr\linewidth-2\fboxsep\relax}
{\small\sffamily\bfseries YuE2 input}\hfill
{\small\sffamily Contemporary pop}\par\smallskip
{\small Generate a melody-only ABC transcription without chord symbols,
then generate music with codec tokens from the given conditions.\par}
\smallskip
{\footnotesize The input lyrics are the four-line verse shown below.}
\end{minipage}}
\par\medskip
\begin{minipage}{\linewidth}
{\small\sffamily\bfseries Generated vocal melody}\hfill
{\footnotesize\sffamily Night bakery}\par\smallskip
\includegraphics[width=\linewidth]{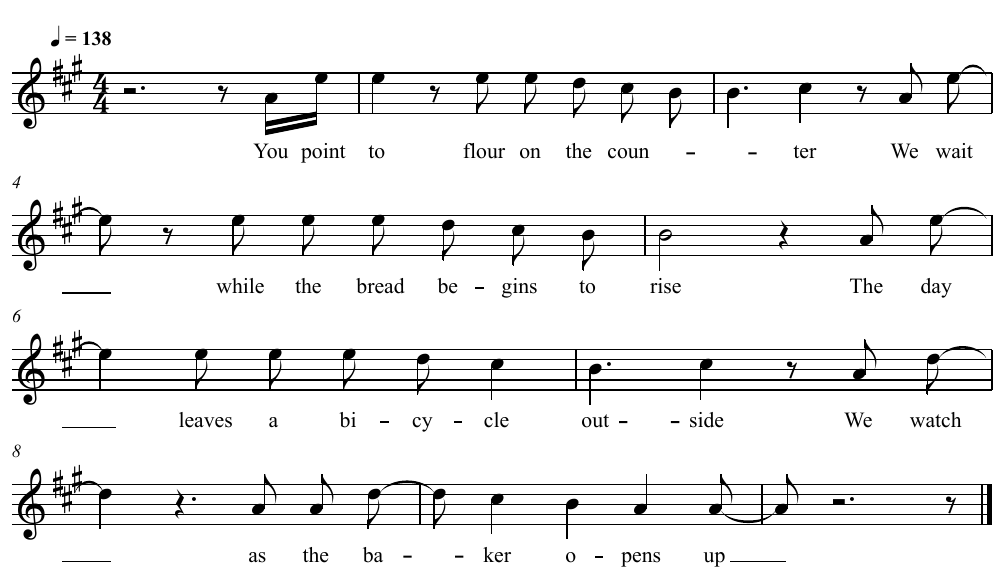}
\par\smallskip
{\color{black!25}\hrule height 0.4pt}\smallskip
{\footnotesize\sffamily Audio: 30.12\,s\hfill
CLAP margin: 0.259\hfill Gemini: contemporary pop}
\end{minipage}
\endgroup
\caption{A data construction example with vocal measures 1--10 from the
generated ABC. Lyric underlay is illustrative, not measured audio alignment.
CLAP and Gemini 3.6 Flash verify the style; all five versions of these lyrics
pass filtering and are retained.}
\label{fig:dataset-example}
\end{figure}

Figure~\ref{fig:dataset-example} illustrates a contemporary-pop sample
generated from the lyrics and style text with score planning enabled
and no input score.

\subsection{Score-Header Distributions Across Styles}
\label{app:data-headers}

Before writing the first note, an ABC score fixes its meter, tempo and key in
the header. Table~\ref{tab:score-headers} compares these settings across styles,
with the scores written for the same lyrics under the empty tag as reference.

\begin{table}[htbp]
\centering
\small
\setlength{\tabcolsep}{4pt}
\setlength{\abovecaptionskip}{0pt}\setlength{\belowcaptionskip}{5pt}
\caption{ABC header fields of the 345 retained samples and of the empty-tag scores for the same 69 lyrics. Minor is the share of scores in a minor key; meter columns count scores per meter; tempo is given in quarter notes per minute.}
\label{tab:score-headers}
\begin{tabular}{lrrrrrll}
\toprule
Group & $n$ & Minor & 4/4 & 2/4 & 6/8 & Tempo, median (IQR) & Most frequent keys \\
\midrule
Contemporary pop & 69 & 9\% & 56 & 13 & 0 & 100 (90--118) & F (14), D (13), A (9) \\
Metal rock & 69 & 100\% & 40 & 29 & 0 & 90 (87--118) & D\#m (51), Em (6), Dm (4) \\
Classic funk & 69 & 96\% & 66 & 3 & 0 & 105 (100--118) & Gm (21), Dm (20), Em (8) \\
Electric blues & 69 & 100\% & 34 & 0 & 35 & 86 (85--88) & Gm (59), Dm (8), D\#m (2) \\
Country & 69 & 1\% & 57 & 12 & 0 & 80 (78--85) & G (20), Bb (15), Eb (14) \\
\midrule
Empty tag & 69 & 54\% & 56 & 11 & 0 & 93 (85--120) & Gm (11), Dm (9), D (8) \\
\bottomrule
\end{tabular}
\end{table}

Metal rock, electric blues and classic funk are almost always in minor keys,
contemporary pop and country almost always in major keys, and only electric
blues uses 6/8 meter. These three settings alone identify 77.1\% of the styles
(chance level 20\%).

\subsection{Swapping Only the Score}
\label{app:data-score-swap}

For each of the 54 representation-analysis lyrics and each style, we generate
two samples that differ only in whether the score is written under the style
tag; all later stages use the empty tag (Table~\ref{tab:abc-swap}).

\begin{table}[htbp]
\centering
\small
\setlength{\abovecaptionskip}{0pt}\setlength{\belowcaptionskip}{5pt}
\caption{Swapping only the score. Both arms share all seeds; the music-token and acoustic stages always receive the empty tag and no intervention. A hit means the target style has the highest CLAP score; the margin is the target score minus the best other score.}
\label{tab:abc-swap}
\begin{tabular}{lrrr}
\toprule
Style & \multicolumn{2}{c}{CLAP hits} & Margin change \\
 & Empty-tag score & Style-tag score & \\
\midrule
Contemporary pop & 48/54 & 53/54 & +0.050 \\
Metal rock & 0/54 & 33/54 & +0.305 \\
Classic funk & 0/54 & 9/54 & +0.228 \\
Electric blues & 2/54 & 34/54 & +0.179 \\
Country & 4/54 & 39/54 & +0.141 \\
\midrule
Non-pop styles & 6/216 & 115/216 & +0.213 \\
\bottomrule
\end{tabular}
\end{table}

Replacing only the score raises the non-pop hits from 6 to 115 of 216, about
81\% of the margin gain obtained by prompting both autoregressive stages
(Appendix~\ref{app:stage-ablation}). The symbolic encoder CLaMP
3~\citep{wu2025clamp3}, judging the scores themselves, agrees overall but not
per style, so we use it only as supporting evidence.

\subsection{Stage-wise Tag Ablation}
\label{app:stage-ablation}

Each of the three stages either keeps the style tag or uses the empty tag,
giving eight conditions. The main effect of a
stage is its average effect when only that stage is switched, over all settings
of the other two. The Shapley value is also an average of the stage's gain over
the settings of the other two stages, but with different weights: the cases in
which neither or both of the other stages carry the tag each receive 1/3, and
the two cases in which exactly one of them does each receive 1/6. These weights
make the three contributions sum exactly to the total gain.

\begin{table}[htbp]
\centering
\small
\setlength{\abovecaptionskip}{0pt}\setlength{\belowcaptionskip}{5pt}
\caption{Effect of each stage in the stage-wise tag ablation (54
representation-analysis lyrics, four non-pop styles, one seed pair).}
\label{tab:stage-ablation}
\begin{tabular}{lrr}
\toprule
Stage & Main effect (margin) & Shapley share \\
\midrule
Score & +0.200 & 76\% \\
Music tokens & +0.061 & 23\% \\
Acoustic rendering & +0.001 & 0\% \\
Full tag, all stages & +0.262 & 100\% \\
\bottomrule
\end{tabular}
\end{table}

\begin{figure}[htbp]
\centering
\begin{minipage}[c]{0.6\linewidth}
\centering
\includegraphics[width=\linewidth]{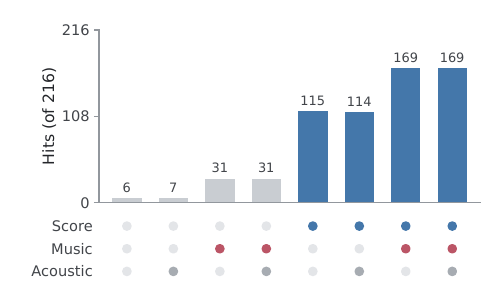}
\end{minipage}\hfill
\begin{minipage}[c]{0.37\linewidth}
\caption{Hits of the four non-pop styles (out of 216) in the eight conditions of
the stage-wise ablation; the dots below mark the stages that keep the style
tag.}
\label{fig:stage-factorial-hits}
\end{minipage}
\end{figure}

The two autoregressive stages overlap slightly: once the score uses the style
tag, adding it to the music-token stage gains about one third less, and acoustic
rendering interacts with neither. Contemporary pop is already hit under the empty
prompt (48 of 54) and is left out.

\FloatBarrier
\section{Implementation Details}
\label{app:implementation}

\subsection{Sublayer Importance}
\label{app:sublayer-importance}

A layer with a large local writing contribution does not necessarily have a large
effect on the final representation. We therefore add a perturbation along the
style direction to one sublayer at a time, recompute the remaining layers, and
take the change of the final representation along the style direction, divided
by the perturbation size, as the importance of that sublayer. In the score stage
we use the KL-weighted direction: the perturbation is added at every score
position, and the readout is averaged over positions with KL weights. We also
record how much of the output KL induced by the style tag the perturbation
removes.

\begin{figure}[htbp]
\centering
\includegraphics[width=\linewidth]{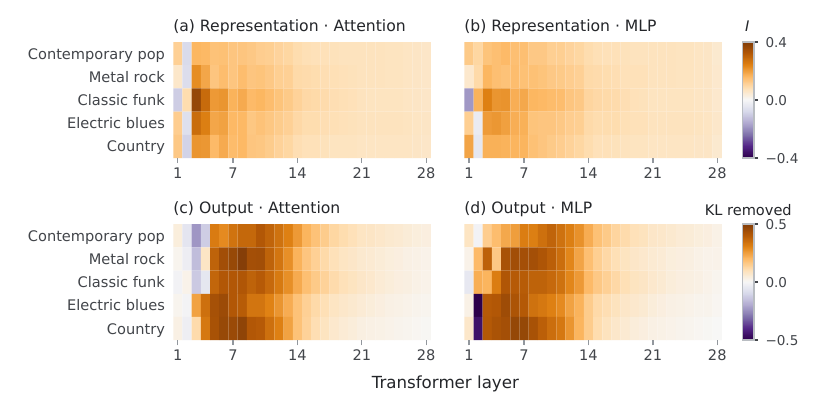}
\caption{Sublayer importance in the score stage (54 representation-analysis
lyrics). Top: change of the final representation. Bottom: fraction of the
output KL removed. Left: attention sublayers; right: MLP sublayers. Color scales
are bounded by the largest positive value; a few more negative cells saturate.}
\label{fig:sublayer-importance-klw}
\end{figure}

Both readouts are largest in the lower and middle layers, roughly layers 3 to 12,
and decrease with depth; after layer 20, perturbing a single sublayer removes
less than a tenth of the output KL. In contrast, the MLP neurons selected by
ranking write scores across all layers lie mostly in the middle and upper
layers, with about a quarter in the last layer. The two quantities measure
different things: the first is the downstream effect of pushing a whole
sublayer along the direction, the second is how much a single component writes.

\paragraph{Budget allocation.} The circuit budget is ten times the number of
sublayers, i.e., 560 components per stage, split 7:3 between MLP neurons and
attention heads. In the music-token stage the budget is allocated to layers in
two steps. First, each sublayer receives a minimum quota, so that lower layers,
whose hidden states remain visible to attention throughout generation, can still
encode style cues. Second, the remaining budget is distributed in proportion to
sublayer importance, emphasizing sublayers with more influence on the final style
representation~\citep{wang2025emotioncircuits}. Within each sublayer we select the
highest-scoring components. In the score stage, the KL-weighted scores are
computed from the actual injected quantity and already reflect how much each
layer writes, so we drop the per-layer quotas and rank all layers together.
Allocating the score-stage budget by sublayer importance may further improve the
circuit, which we leave to future work.

\begin{figure}[htbp]
\centering
\includegraphics[width=\linewidth]{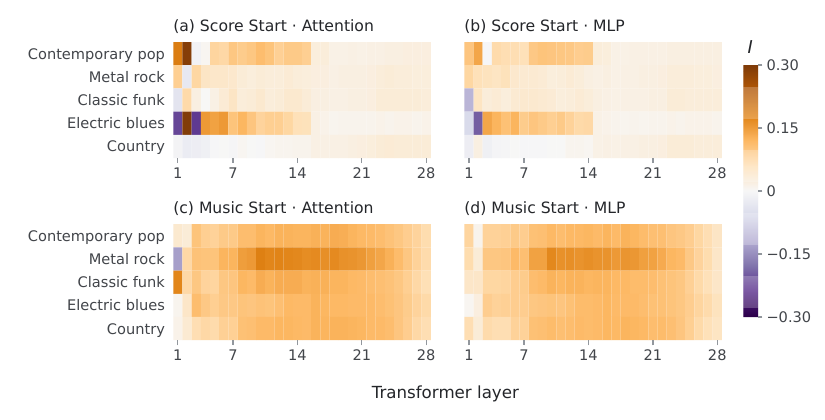}
\caption{Sublayer importance of the single-position directions. Top: the score
start marker (Score Start); bottom: Music Start. Left: attention sublayers;
right: MLP sublayers. Positive values promote the target style and negative
values work against it.}
\label{fig:sublayer-importance}
\end{figure}

The music-token stage allocates per-layer quotas by the sublayer importance at
Music Start (Figure~\ref{fig:sublayer-importance}, bottom row): each layer
receives at least 4 neurons and 2 heads, the rest of the budget is distributed
in proportion to the positive part of the importance with largest-remainder
rounding, and the extra quota of any layer is capped at 40\% of the budget for
that component type.

\FloatBarrier
\subsection{Random Component Controls}
\label{app:random-controls}

Random controls draw, in each layer, the same number of components as the
target, and are averaged over three random seeds (ten seeds for the
single-position experiments in Appendix~\ref{app:single-circuit}). Each layer has
only 16 attention heads, and the target selects about a third of them in most
layers, so a random draw that allowed overlap would share many heads with the
target. We therefore draw only from components that the target does not select;
where a layer has too few left, the shortfall moves to the nearest layer that
also contains target components. For enhancement, a second random control
rescales the injected difference to the target's per-layer norm, which rules out
the injection size itself as an explanation.

\begin{figure}[htbp]
\centering
\includegraphics[width=\linewidth]{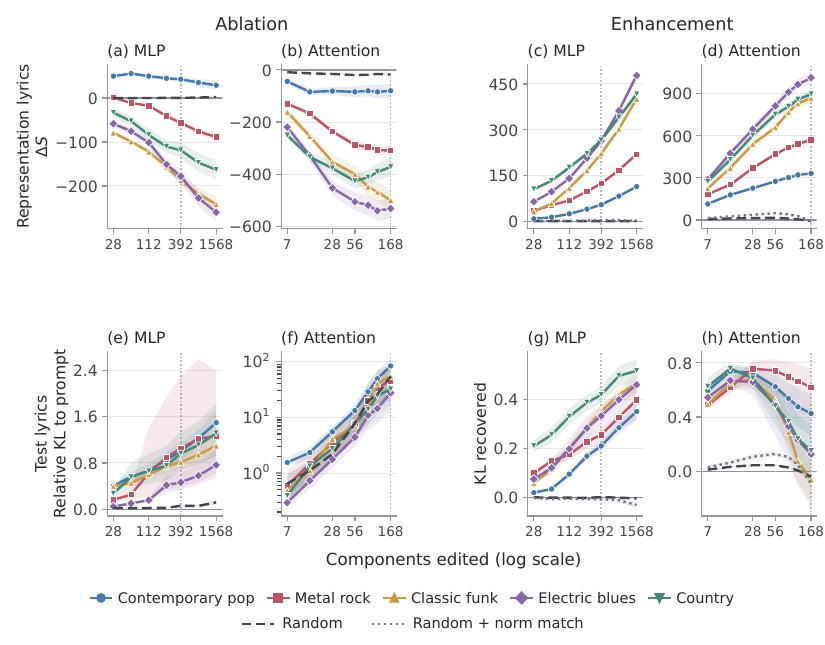}
\caption{Component interventions with the KL-weighted direction as reference.
Top (a--d): style projection on the 54 representation-analysis lyrics. Bottom
(e--h): change of the output distribution on the test set. Columns as in
Figure~\ref{fig:klw-causal}: MLP ablation, attention-head ablation, MLP
enhancement and attention-head enhancement; gray curves are the random
controls.}
\label{fig:klw-causal-full}
\end{figure}

Figure~\ref{fig:klw-causal-full} shows the style projection on the
representation analysis set (top) and the change of the output distribution on
the test set (bottom). For ablation, the bottom row gives the KL between the
intervened output and the style-tag output, divided by the KL between the
empty-tag output and the style-tag output; for enhancement, it gives one
minus the same ratio, i.e., the fraction recovered.

The MLP results agree on the two sets, and random neurons have almost no effect.
If positions are averaged equally rather than with KL weights, the target
components still clearly beat the random ones, but the effect of MLP ablation
becomes much smaller, which shows that zeroing these neurons matters mainly at
positions with high KL. Attention heads are also far stronger than random
components on the style projection (panels b and d). Output-level ablation is the
exception: zeroing the same number of random heads also changes the output
substantially (panel f, log scale), so the output effect of zeroing heads is not
specific. For enhancement, the recovered fraction is highest with 14 to 28 heads
and then decreases, while random heads stay near zero at every budget (panel h).

\subsection{KL-weighted Direction and Component Selection}
\label{app:klw-details}

The KL weights and directions use only the representation analysis set. For each
lyric group, one score written under the empty tag serves as the shared
history, and we read the next-token distribution and the component activations
at every score position under the empty tag and under each of the five style
tags. Within each lyric group, the KL values are normalized into weights, the
activation differences are summed with these weights, and the result is averaged
over lyrics. The same weighted average of the residual stream gives
$\rho_s^{(\ell)}$, and $u_s^{(\ell)}=\rho_s^{(\ell)}/\|\rho_s^{(\ell)}\|_2$, taken
after the MLP for neurons and after attention for heads. The component scores are
\begin{equation}
\begin{gathered}
c_{s,j}^{(\ell)}=\delta_{s,j}^{(\ell)}\,\big\langle W_{\mathrm{down},:,j}^{(\ell)},\,u_s^{(\ell)}\big\rangle ,\\
c_{s,h}^{(\ell)}=\big\langle W_{O,h}^{(\ell)}\,\delta_{s,h}^{(\ell)},\,u_s^{(\ell)}\big\rangle ,
\end{gathered}
\end{equation}
where $\delta_{s,j}^{(\ell)}$ and $\delta_{s,h}^{(\ell)}$ are the components of
the KL-weighted difference on neuron $j$ and head $h$, and $W_{O,h}^{(\ell)}$ is
the column block of the output projection for that head. MLP neurons and
attention heads are ranked separately, and the 392 neurons and 168 heads with the
highest positive scores are kept.

In the causal validation, ablation sets $a_J\leftarrow 0$ and enhancement sets
$a_J\leftarrow a_J+\delta_{s,J}$, both at every score position. The readout is
\begin{equation}
\begin{gathered}
\Delta S_{i,s}=\sum_{p} w_{i,s,p}\sum_{\ell=1}^{28}\big\langle \widetilde h_{i,p}^{(\ell)}-h_{i,p}^{(\ell)},\,u_s^{(\ell)}\big\rangle ,\\
w_{i,s,p}=\frac{\mathrm{KL}_{i,s,p}}{\sum_{p'}\mathrm{KL}_{i,s,p'}} ,
\end{gathered}
\end{equation}
where $\widetilde h$ is the residual after the intervention, read after each
sublayer and projected on the $u_s^{(\ell)}$ of that sublayer output. At the
output, we
also record the KL between the intervened output and the style-tag output,
relative to the KL between the empty-tag and style-tag outputs.

\subsection{Injection Strength}
\label{app:dose}

The KL-weighted direction uses the raw difference, so the injection strength must
be limited (Table~\ref{tab:dose}).

\begin{table}[htbp]
\centering
\small
\setlength{\abovecaptionskip}{0pt}\setlength{\belowcaptionskip}{5pt}
\caption{Injection strength on four representation-analysis lyrics and the four non-pop styles. At $\alpha=2.0$ all score headers are
corrupted, so its hits cannot be read as style control.}
\label{tab:dose}
\begin{tabular}{rrrrr}
\toprule
Strength $\alpha$ & CLAP hits & Complete & Mean ABC tokens & Mean audio (s) \\
\midrule
0.5 & 14/16 & 16/16 & 697 & 83 \\
1.0 & 13/16 & 13/16 & 1378 & 143 \\
2.0 & 11/16 & 13/16 & 915 & 175 \\
\bottomrule
\end{tabular}
\end{table}

From $\alpha=1.0$ on, scores become much longer and some are truncated, so we use
$\alpha=0.5$. The budget is 392 MLP neurons (0.23\% of all neurons) and 168
attention heads per stage and style.

\paragraph{Music-token stage.} The KL-weighted, position-dependent aggregation is
applied only to the score stage, because the stage localization of
\S\ref{sec:exp-stage} shows that the score stage carries most of the style
effect and is therefore the stage worth modeling in detail; the music-token
stage, which contributes about a quarter, is steered with the simpler
equal-weight mean of the first 128 positions, with strength 1.0 (1.5 for classic
funk and electric blues). This music-token intervention is shared by all circuit
conditions and by CAA, so the conditions of Table~\ref{tab:steering} differ only
in the score stage. Injecting throughout the stage rather than only at its first
steps follows the same reasoning as in the score stage: style decisions are
spread over the whole sequence.

\FloatBarrier
\section{Extended Analyses}
\label{app:analyses}

\subsection{Single-position Style Directions}
\label{app:single-direction}

This section shows the effect of within-lyric centering for the single-position
directions of \S\ref{sec:single-position}, the style separation of component
activations, and the representations at more layers, using all 345 samples and
t-SNE~\citep{vandermaaten2008tsne}. Each autoregressive stage contributes three
positions: its first position, the mean of its middle 64 tokens and its end
marker.

\begin{figure}[htbp]
\centering
\includegraphics[width=0.85\linewidth]{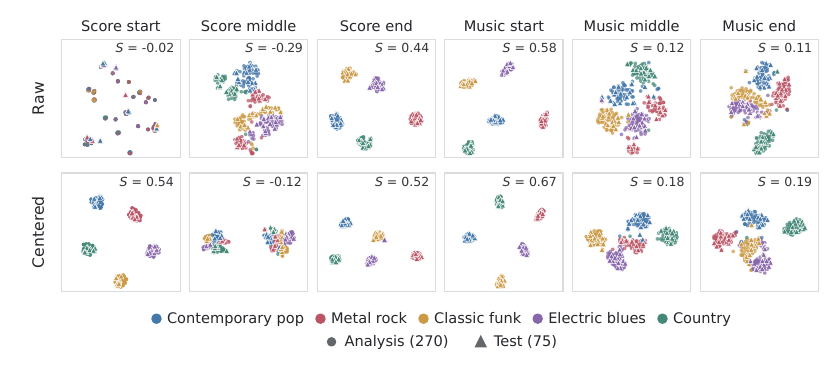}
\caption{Hidden states at six positions of layer 14 (t-SNE). Each stage
contributes its first position, the mean of its middle 64 tokens and its end
marker. Top: raw hidden states; bottom: style representations after within-lyric
centering. Circles are representation-analysis samples and triangles are test
samples; $S$ is the cosine silhouette score in the original space.}
\label{fig:style-representations}
\end{figure}

After centering, the five styles separate clearly at the start and end of the
score and at the start of the music tokens, and the test samples fall into the
clusters of their styles; in the middle of the score, styles remain hard to
distinguish even after centering.

\begin{figure}[htbp]
\centering
\includegraphics[width=0.85\linewidth]{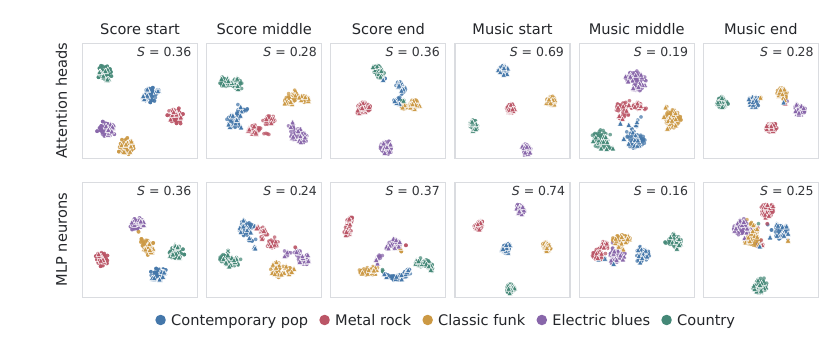}
\caption{Attention-head outputs (top) and MLP-neuron activations (bottom) of the
five styles at six positions of layer 14 (t-SNE, after within-lyric centering).
Markers and $S$ as in Figure~\ref{fig:style-representations}.}
\label{fig:style-sites-components}
\end{figure}

The two component activations where circuits live also separate the styles. In
the middle of the score, hidden states barely distinguish styles ($S=-0.12$),
whereas attention-head outputs and MLP-neuron activations still do ($S=0.28$ and
$0.24$).

\begin{figure}[htbp]
\centering
\includegraphics[width=0.85\linewidth]{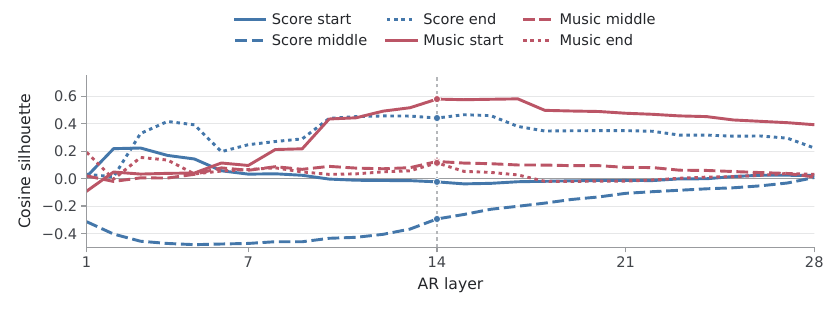}
\caption{Style separation of raw hidden states across layers at six positions
(cosine silhouette score). Blue: score stage; red: music-token stage. Solid,
dashed and dotted lines are the start, middle and end.}
\label{fig:style-separation-layers}
\end{figure}

\begin{figure}[htbp]
\centering
\includegraphics[width=0.88\linewidth]{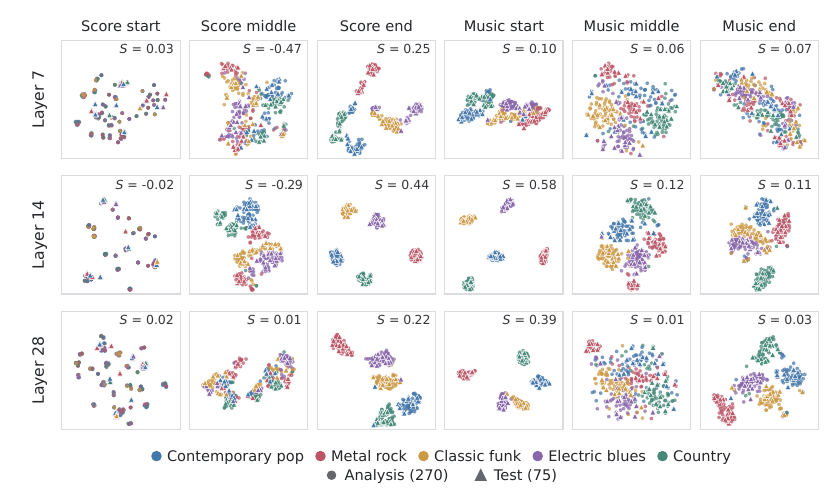}\\[4pt]
\includegraphics[width=0.88\linewidth]{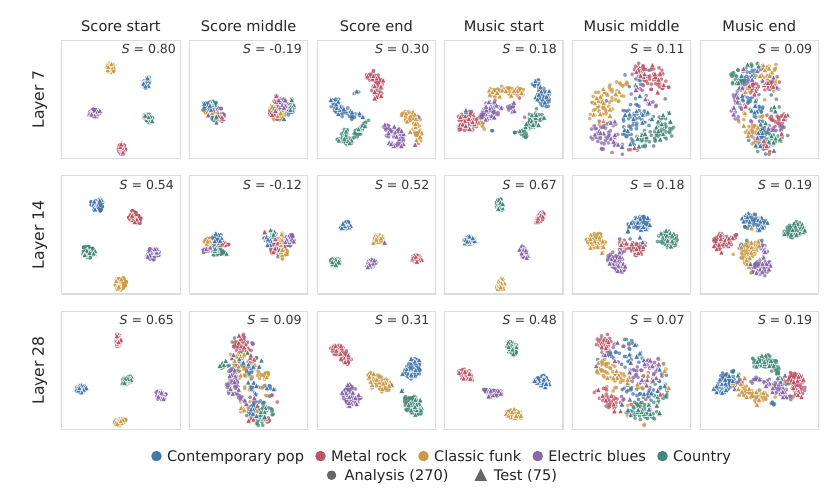}
\caption{Hidden states at six positions in layers 7, 14 and 28. Top: raw hidden
states, with $S$ the cosine silhouette score in the original space. Bottom: the
same after within-lyric centering.}
\label{fig:raw-layer-overview}
\end{figure}

\subsection{Single-position Circuits and Causal Validation}
\label{app:single-circuit}

\begin{figure}[htbp]
\centering
\includegraphics[width=0.92\linewidth]{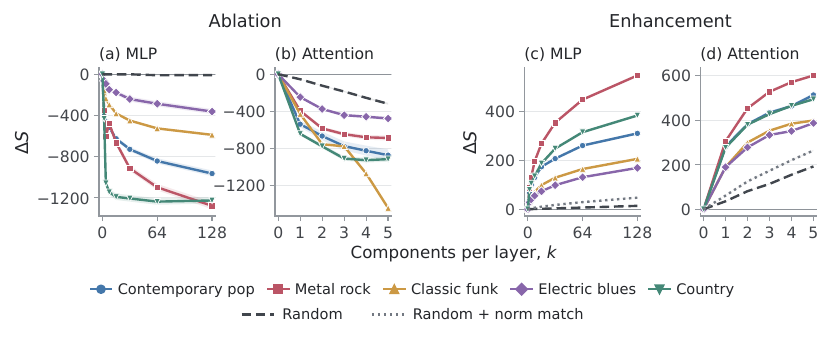}\\[3pt]
\includegraphics[width=0.92\linewidth]{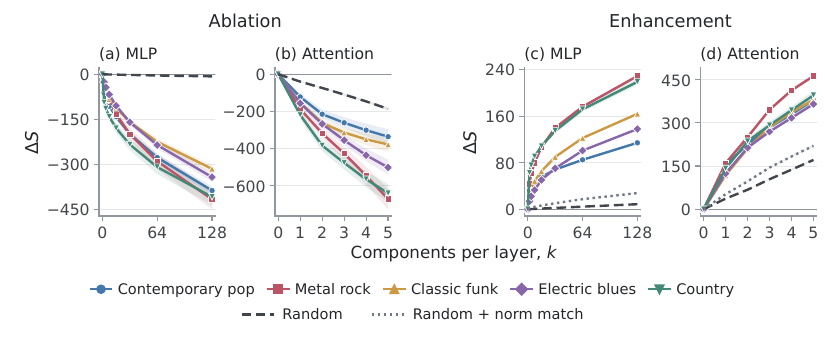}\\[3pt]
\includegraphics[width=0.92\linewidth]{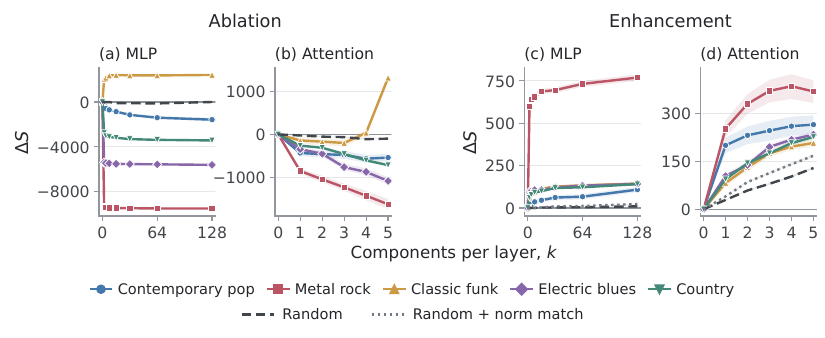}
\caption{Component interventions at Music Start (top), Music Middle (middle) and
Music End (bottom). In each row, from left to right: MLP ablation, attention-head
ablation, MLP enhancement and attention-head enhancement.}
\label{fig:circuit-music-start}
\end{figure}

This section reports the single-position circuits of \S\ref{sec:single-position}
at three positions of the music-token stage; the definitions are given in
\S\ref{sec:single-position}. Figures~\ref{fig:circuit-music-start}
to~\ref{fig:circuit-music-start} use the direction at the start of the music tokens
(Music Start) and repeat the experiment at two further positions: Music Middle,
the middle 64 tokens of the music sequence, and Music End, the position after
the music end marker. The horizontal axis is the number of edited components per
layer.

At all three positions, removing a few high-contribution components clearly
weakens the style projection, and injecting the difference pushes the
representation toward the target style without a style prompt; both effects are
clearly stronger than for random components of the same number and injection norm. Style
at a single position is thus concentrated in a small number of components. Music
Middle agrees with Music Start, while Music End is dominated by a few components
and varies more across styles.

\FloatBarrier
\subsection{Cross-position Stability}
\label{app:cross-position}

We extract style directions at positions where 0, 32, 64, 128 and 256 tokens have
been generated in each of the two autoregressive stages, and select components
independently at each position (Table~\ref{tab:cross-position}).

\begin{table}[htbp]
\centering
\small
\setlength{\abovecaptionskip}{0pt}\setlength{\belowcaptionskip}{5pt}
\caption{Stability of style readouts and component selection (layer 28). A readout is a cosine-prototype style classifier fitted on the
representation-analysis set; ``above chance'' means that its test accuracy beats
a family-coherent label-permutation null (Holm-adjusted $p<0.05$). One
representation-analysis lyric has a score shorter than 256 tokens and no step
256, so directions at all positions use the other 53; style readouts are
evaluated on 13 test lyrics.}
\label{tab:cross-position}
\begin{tabular}{p{0.44\linewidth}p{0.48\linewidth}}
\toprule
Comparison & Result \\
\midrule
Style readout trained and tested at the same position & above chance in 79 of 80 cases \\
Style readout transferred across positions & above chance in 228 of 320 cases; 21.5\% when trained at the score start and tested at step 256 \\
MLP neurons reselected after resampling lyrics (score start) & Jaccard 0.80--0.89 \\
\bottomrule
\end{tabular}
\end{table}

\begin{figure}[htbp]
\centering
\includegraphics[width=\linewidth]{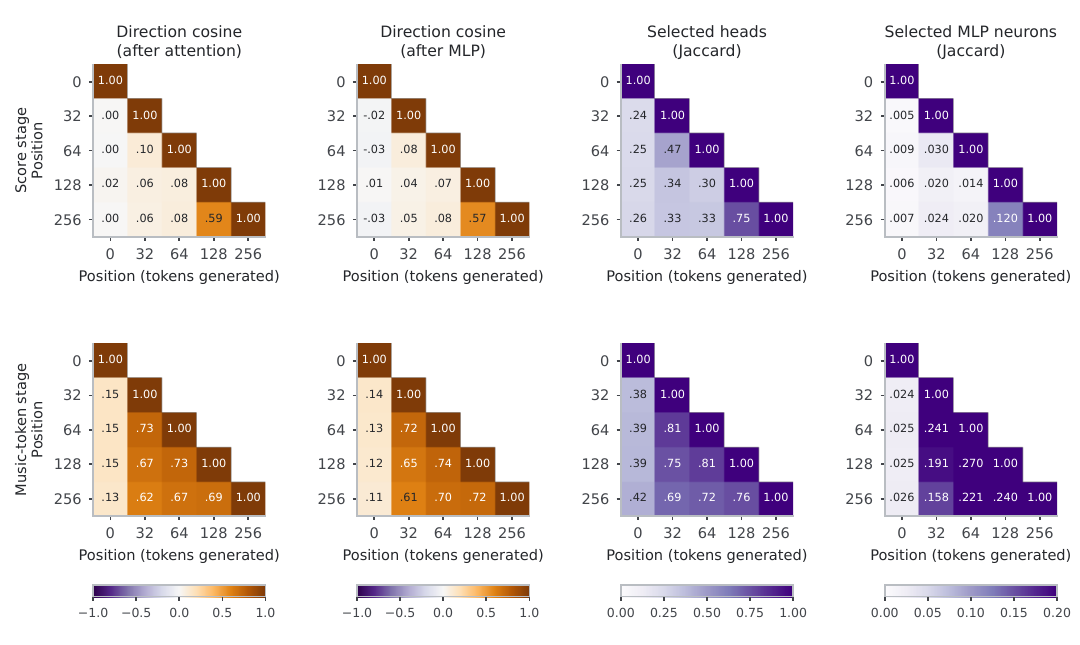}
\caption{Direction cosines and component overlaps between positions in the two
autoregressive stages (top: score; bottom: music tokens), read as
Figure~\ref{fig:cross-position}; values are means over the five styles.}
\label{fig:cross-position-full}
\end{figure}

In the score stage, only steps 128 and 256 are close to each other; in the
music-token stage, the directions and attention heads of all positions after the
start are fairly consistent. KL is also highly concentrated: over the 54
representation-analysis lyrics and five styles, the 5\% of positions with the
highest KL carry 59\% of the total KL on average (median 59\%, minimum 40\%).
A token-type count on six of these lyrics shows that nearly half of the weight
of an equal average over the first 128 steps falls on header format text that
the style barely affects.

\FloatBarrier
\subsection{Robustness of the Steering Results}
\label{app:score-diagnostics}

\paragraph{Paired comparison.} On the test set, paired by lyric, style and seed,
the circuit's CLAP margin matches CAA's and exceeds that of every other
condition (lyric-family bootstrap intervals above zero).

\paragraph{Scores and failure modes.} CLaMP 3 judges the circuit's metal-rock and
country scores about as often as the prompt's to be in the target style, but
rarely its funk and blues scores; it seldom recognizes blues scores even under
the prompt. When the style prompt misses, CLAP mostly hears contemporary pop,
the model's default style; when the circuit misses, it mostly hears electric
blues.

\paragraph{Length.} CLAP averages 10-second windows over the whole track, so we
recompute the hits on the first 30 seconds and on the middle 20 seconds
(Table~\ref{tab:fixed-length}).

\begin{table}[htbp]
\centering
\small
\setlength{\abovecaptionskip}{0pt}\setlength{\belowcaptionskip}{5pt}
\caption{CLAP hits on fixed-length crops (test set, 150 generations per
condition).}
\label{tab:fixed-length}
\begin{tabular}{lrrrr}
\toprule
Condition & Mean audio (s) & Whole track & First 30 s & Middle 20 s \\
\midrule
Empty-tag baseline & 53 & 30 & 30 & 30 \\
Style prompt (all stages) & 66 & 113 & 123 & 93 \\
Single-position circuit & 64 & 73 & 73 & 62 \\
Mean-direction circuit & 72 & 90 & 91 & 89 \\
Random circuit & 58 & 34 & 37 & 28 \\
Dynamic circuit & 84 & 133 & 122 & 126 \\
\bottomrule
\end{tabular}
\end{table}

With equal crops, the circuit matches the prompt on the first 30 seconds and
leads on the middle 20: the prompt's style is concentrated at the start, while the
circuit keeps it through the song, so its advantage does not come from length.

\FloatBarrier
\section{Supplementary Evaluation of Style Steering}
\label{app:audio}

\subsection{Per-style Results and Further Baselines}
\label{app:steering-extra}

Table~\ref{tab:clap-per-style} gives the CLAP hits per style of all conditions on
the test set. The two prompts write the same score and music tokens and differ
only in acoustic rendering, so their hits are identical; the main text keeps the
all-stage prompt.

\begin{table}[htbp]
\centering
\small
\setlength{\tabcolsep}{5pt}
\setlength{\abovecaptionskip}{0pt}\setlength{\belowcaptionskip}{5pt}
\caption{CLAP hits on the 15 test lyrics with two seed pairs.}
\label{tab:clap-per-style}
\begin{tabular}{lrrrrrr}
\toprule
Condition & Pop & Metal & Funk & Blues & Country & Total \\
\midrule
Empty-tag baseline & 27/30 & 0/30 & 0/30 & 0/30 & 3/30 & 30/150 \\
Style prompt (AR stages) & 29/30 & 17/30 & 13/30 & 29/30 & 25/30 & 113/150 \\
Style prompt (all stages) & 29/30 & 17/30 & 13/30 & 29/30 & 25/30 & 113/150 \\
Single-position circuit & 30/30 & 5/30 & 2/30 & 15/30 & 21/30 & 73/150 \\
Mean-direction circuit & 30/30 & 10/30 & 13/30 & 17/30 & 20/30 & 90/150 \\
Random circuit & 27/30 & 0/30 & 0/30 & 3/30 & 4/30 & 34/150 \\
CAA & 26/30 & 22/30 & 26/30 & 21/30 & 23/30 & 118/150 \\
Dynamic circuit & 30/30 & 24/30 & 25/30 & 27/30 & 27/30 & 133/150 \\
\bottomrule
\end{tabular}
\end{table}

The single-position circuit uses the direction and components of
\S\ref{sec:single-position} at the first score position (Score Start), injected
at every score step; the mean-direction circuit averages the direction over the
first 128 score steps with equal weights. Both share the component budget,
strength and music-token intervention of the dynamic circuit. They barely work on
metal rock and classic funk and hit far less often in total, so aggregating the
directions with KL weights is the key. CAA matches the circuit only on classic
funk, and most of its misses are truncations.

\paragraph{Conventional steering.} Besides CAA~\citep{rimsky2024caa}, we test two
steering methods without a sparse circuit: the KL-weighted residual direction,
injected like CAA, and dense injection of the KL-weighted difference into all MLP
neurons and attention heads. All keep the style text empty, change only the score-stage intervention, and
have their layer and strength chosen on the same representation-analysis lyrics
as the circuit. The two were
run on 70 test generations, on which Table~\ref{tab:conventional} also shows CAA.

\begin{table}[htbp]
\centering
\small
\setlength{\abovecaptionskip}{0pt}\setlength{\belowcaptionskip}{5pt}
\caption{Conventional steering on 70 test generations (14 lyrics and five
styles, first seed pair).}
\label{tab:conventional}
\begin{tabular}{lrrrr}
\toprule
Condition & CLAP & Gemini & Truncated & Length (s) \\
\midrule
Style prompt (all stages) & 56 & 54 & 0 & 61 \\
CAA & 52 & 54 & 12 & 185 \\
KL-weighted residual direction & 60 & 50 & 1 & 65 \\
Dense injection & 56 & 51 & 4 & 91 \\
Dynamic circuit & 65 & 58 & 0 & 73 \\
\bottomrule
\end{tabular}
\end{table}

All three steer the style without style text but hit less often than the
dynamic circuit, and CAA and dense injection often derail generation. The
KL-weighted residual direction uses the same position weights as the circuit
but injects into the residual stream instead of sparse components; it is close
to the circuit under CLAP and lower under Gemini.

\subsection{Gemini 3.6 Flash as a Second Judge}
\label{app:gemini}

Gemini 3.6 Flash~\citep{googledeepmind2026gemini36flash} judges every test-set
generation blind, with the protocol used to build the dataset
(Appendix~\ref{app:data}): the full audio is uploaded under an anonymous file
name, and Gemini chooses among the five styles and a distractor, Chinese
traditional, in an order shuffled for each sample, at temperature 0. The
empty-tag baseline has one audio per lyric and seed, scored against every target
style (Table~\ref{tab:gemini}).

\begin{table}[htbp]
\centering
\small
\setlength{\tabcolsep}{5pt}
\setlength{\abovecaptionskip}{0pt}\setlength{\belowcaptionskip}{5pt}
\caption{Gemini 3.6 Flash hits on the 15 test lyrics with two seed pairs.}
\label{tab:gemini}
\begin{tabular}{lrrrrrr}
\toprule
Condition & Pop & Metal & Funk & Blues & Country & Total \\
\midrule
Empty-tag baseline & 14/30 & 1/30 & 0/30 & 0/30 & 15/30 & 30/150 \\
Style prompt (all stages) & 24/30 & 12/30 & 15/30 & 30/30 & 30/30 & 111/150 \\
Single-position circuit & 25/30 & 13/30 & 7/30 & 9/30 & 30/30 & 84/150 \\
Mean-direction circuit & 14/30 & 13/30 & 11/30 & 15/30 & 26/30 & 79/150 \\
Random circuit & 13/30 & 0/30 & 0/30 & 0/30 & 20/30 & 33/150 \\
CAA & 24/30 & 22/30 & 25/30 & 23/30 & 26/30 & 120/150 \\
Dynamic circuit & 25/30 & 23/30 & 22/30 & 27/30 & 29/30 & 126/150 \\
\bottomrule
\end{tabular}
\end{table}

Pairing the same lyric, style and seed, the dynamic circuit beats the style
prompt under both judges (exact McNemar test; CLAP: 32 against 12 discordant
pairs, $p=0.004$; Gemini: 31 against 16, $p=0.04$). It beats CAA under CLAP (28
against 13, $p=0.03$) but not significantly under Gemini (25 against 19,
$p=0.45$). Since Gemini was also used to filter the dataset, it is not a fully
independent judge.

\subsection{Subjective Evaluation}
\label{app:listening}

We draw 20-second clips of the empty-tag baseline, the style prompt and the
dynamic circuit from the test generations, loudness-normalized and anonymized.
In each question, a listener sees the target style, rates the three clips of one
lyric and seed for style fit from 1 (not at all) to 5 (very much), and picks the
closest; the order of the conditions rotates. Before the questions, listeners
hear short examples of each style taken from style-prompt generations, which if
anything favors the prompt. A listener who rates the baseline 4 or higher on
average for metal rock, classic funk and electric blues, which it almost never
matches, is excluded (a rule fixed in advance), and only completed questions
count.

Table~\ref{tab:listening} gives the preliminary results (eleven listeners, 54
completed questions, 43 on the non-pop styles). On the non-pop questions both
the circuit and the prompt are rated far above the baseline (Wilcoxon
signed-rank test, $p<0.001$ for each); between the two, the prompt is rated
higher in 19 questions and the circuit in 13, with 11 ties ($p=0.64$), and the
prompt is picked as closest more often (27 against 14, binomial $p=0.06$). On
pop, the circuit is rated highest. The sample cannot yet separate the circuit
from the prompt.

\begin{table}[htbp]
\centering
\small
\setlength{\tabcolsep}{5pt}
\setlength{\abovecaptionskip}{0pt}\setlength{\belowcaptionskip}{5pt}
\caption{Mean style-fit ratings (1--5) and the number of non-pop questions in
which each condition was picked as closest.}
\label{tab:listening}
\begin{tabular}{lrrrrrrr}
\toprule
Condition & Pop & Metal & Funk & Blues & Country & Non-pop & Picked \\
\midrule
Empty-tag baseline & 3.64 & 1.09 & 1.45 & 1.70 & 2.55 & 1.70 & 2/43 \\
Style prompt & 4.00 & 3.82 & 4.09 & 3.40 & 3.82 & 3.79 & 27/43 \\
Dynamic circuit & 4.36 & 3.73 & 3.18 & 3.70 & 3.91 & 3.63 & 14/43 \\
\bottomrule
\end{tabular}
\end{table}

\subsection{Score Example and Instrumentation Measurement}
\label{app:instrumentation}

\begin{figure}[htbp]
\centering
\includegraphics[width=\linewidth]{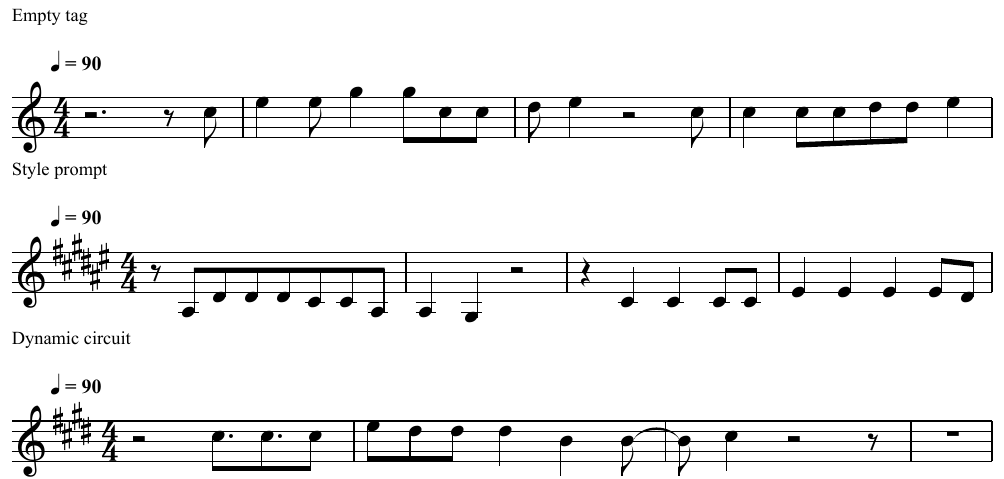}
\caption{Vocal line of one test lyric with target style metal rock under the
three conditions, first four bars after the vocal entry.}
\label{fig:score-exemplar}
\end{figure}

Figure~\ref{fig:score-exemplar} shows the vocal line of one test lyric under the
three conditions: the empty tag writes in C major, the style prompt moves to
D$\sharp$m, and the circuit writes a narrower, smoother melody in C$\sharp$m.

\paragraph{Key of the empty-tag scores.} The empty-tag row of
Figure~\ref{fig:score-change} pools the two seed pairs of the test set, and its
key distribution is set by the seed rather than by the lyrics: the score header
is sampled within the first few tokens, and with the same seed nearly every
lyric receives the same key (12 of 14 scores in C major under the first seed
pair; E$\flat$ and B$\flat$ under the second; a spread of G minor, D minor and D major for
the 69 empty-tag scores of Appendix~\ref{app:data-headers}, written with a third
seed). The empty tag therefore has no style-typical key, whereas the style prompt
and the circuit place the scores in the style's typical key under both seed
pairs.

The score carries no instruments, so instrumentation is decided in the later
stages. With the CLAP model used for style, we score each test audio against 13
instrument descriptions (``This audio features \{instrument\}.'';
Table~\ref{tab:instruments}) and subtract the score of the empty-tag audio of the
same lyric and seed.

\begin{table}[htbp]
\centering
\small
\setlength{\abovecaptionskip}{0pt}\setlength{\belowcaptionskip}{5pt}
\caption{Instrument tags of the instrumentation analysis.}
\label{tab:instruments}
\begin{tabular}{llll}
\toprule
Family & Label & Instrument in the description & Typical of \\
\midrule
Guitars & distorted & distorted electric guitar & metal rock, electric blues \\
 & clean el. & clean electric guitar & classic funk, electric blues \\
 & acoustic & acoustic guitar & country, contemporary pop \\
 & pedal steel & pedal steel guitar & country \\
Rhythm & drum kit & a loud drum kit & metal rock, classic funk \\
 & slap bass & slap bass & classic funk \\
Keys & organ & a Hammond organ & electric blues \\
 & piano & piano & contemporary pop \\
 & synth & synthesizer & contemporary pop \\
Folk & banjo & banjo & country \\
 & fiddle & fiddle & country \\
 & harmonica & harmonica & electric blues, country \\
Brass & brass & a brass section with horns & classic funk \\
\bottomrule
\end{tabular}
\end{table}

Empty-tag audio is dominated by acoustic guitar and piano. CLAP hears electric
blues as distorted guitar with drums, close to metal rock, which matches its
confusion of the two styles.

\subsection{Content and Melody Preservation}
\label{app:content}

Every condition starts from the lyrics and writes a new score, so only the
lyrics are fixed. We compare each score with the empty-tag score of the same
lyric and seed, by CLaMP 3 score features and by the similarity of the vocal
line's intervals and rhythm, with the empty-tag score of the other seed pair
(resampling) as the reference for natural variation (Table~\ref{tab:content}).

\begin{table}[htbp]
\centering
\small
\setlength{\abovecaptionskip}{0pt}\setlength{\belowcaptionskip}{5pt}
\caption{Closeness of each condition's scores to the empty-tag score of the same
lyric and seed (test set).}
\label{tab:content}
\begin{tabular}{lrrrr}
\toprule
Condition & CLaMP 3 cosine & Lyric advantage & Interval sim. & Rhythm sim. \\
\midrule
Empty-tag resampling (reference) & 0.553 & 0.027 & 0.310 & 0.383 \\
Style prompt & 0.528 & 0.008 & 0.322 & 0.423 \\
Dynamic circuit & 0.561 & 0.022 & 0.323 & 0.436 \\
Mean-direction circuit & 0.678 & 0.099 & 0.397 & 0.508 \\
Single-position circuit & 0.645 & 0.064 & 0.361 & 0.499 \\
Random circuit & 0.747 & 0.148 & 0.483 & 0.586 \\
\bottomrule
\end{tabular}
\end{table}

The circuit and the prompt both stay at the level of resampling: starting from
the lyrics, both write a new melody. Conditions closer to the empty-tag score hit
the target style less often, consistent with the score being the main carrier of
style (Appendix~\ref{app:data-score-swap}). Steering a fixed score is left to
future work.

\section{Extension to Vocal Gender}
\label{app:vocal-gender}

To test whether the circuit picture found for genre extends to another style
attribute, we repeat the analysis for vocal gender, using the descriptors ``male vocal'' and ``female vocal'' as the two conditions;
the gender descriptor plays the role of the style tag and an empty descriptor
that of the empty tag. We use 128 short lyrics from 32 lyric families, split by family into 96 representation-analysis and 32 evaluation lyrics. Directions, component selections, and intervention settings are estimated on the representation-analysis split only. Unlike the genre experiments, each lyric uses one neutral ABC score that is reused byte-identically across the two conditions. The experiment therefore isolates how vocal-gender conditioning acts after the score has been fixed. For representation comparisons, both forward passes also replay the same generated music-token history and differ only in the gender descriptor; results are averaged over male-generated and female-generated history arms.

Three findings emerge. With the score fixed, the remaining vocal-gender effect is carried almost entirely by the music-token stage rather than by acoustic rendering. Within the music-token stage, Music Start differs sharply from later positions, which are substantially more stable. Finally, the selected components can steer the generated vocal gender without gender text, although averaging over the first 128 positions does not improve on the Music-Start circuit.

\subsection{Post-Score Vocal-Gender Control Is Mainly Autoregressive}
\label{app:vocal-stage}

We isolate the two post-score sources of vocal-gender control by switching the
gender descriptor independently in the music-token stage and in acoustic
rendering, as in the stage-wise ablation of \S\ref{sec:stage-ablation}. Across
42 pairs from 24 lyric families, switching the music tokens changes the CLAP
female-minus-male margin by $+0.123$ on average, compared with only $+0.0018$
for switching the acoustic-rendering condition. The music-token effect is
positive in all 42 pairs and about 68 times larger than the acoustic-rendering
effect, which is positive in 41 of 42 pairs; their interaction is near zero
($+0.0006$). Thus, once the score is fixed, nearly all of the remaining
vocal-gender effect is established in the music-token stage. The contribution of score planning itself is not tested by this design.

\subsection{Vocal-Gender Representations Stabilize after Music Start}
\label{app:vocal-positions}

We next ask whether the same vocal-gender representation persists as music-token generation proceeds. At positions $p\in\{0,32,64,128,256\}$, where $p=0$ is Music Start, we estimate a female-minus-male direction from the shared-history contrast
$h(c_F,x_{<p})-h(c_M,x_{<p})$ and independently select the MLP neurons and attention heads that write along it. We compare positions by direction cosine and component Jaccard overlap, the same quantities used for genre in Figure~\ref{fig:cross-position}. The two history arms give closely aligned directions (cosine $0.92$--$0.99$ after Music Start), so the result is not tied to which generated history is replayed. We report layers 8--28, where the contrast is not dominated by shallow lexical traces of the descriptor tokens.

\begin{figure}[tbp]
    \centering
    \includegraphics[width=0.92\linewidth]{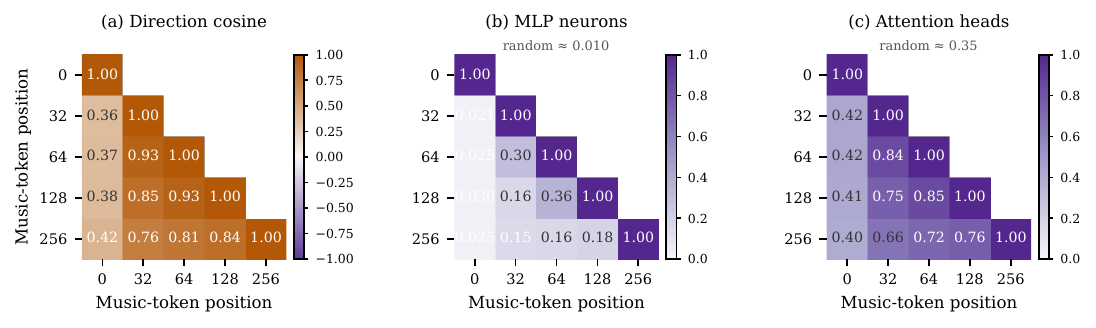}
    \caption{\textbf{Vocal-gender representation across music-token positions.}
    Shared-history female-minus-male contrasts over layers 8--28.
    (a) Cosine between residual directions after the MLP;
    (b) Jaccard overlap of selected MLP neurons;
    (c) Jaccard overlap of selected attention heads.
    Music Start is distinct, while positions 32--256 share a substantially more stable representation.}
    \label{fig:vocal-position}
\end{figure}

The result is a clear transition from Music Start to a more stable later regime (Figure~\ref{fig:vocal-position}). The Music-Start direction has only $0.36$--$0.42$ cosine similarity with positions 32--256, whereas pairwise similarity among the later positions rises to $0.76$--$0.93$. Component overlap changes in the same way. MLP-neuron Jaccard overlap is only $0.02$--$0.03$ between Music Start and later positions but $0.15$--$0.36$ among later positions, compared with a chance level of about $0.01$. Attention-head overlap rises from $0.40$--$0.42$ to $0.66$--$0.85$, compared with chance around $0.35$. Vocal gender also remains readable on held-out lyrics at every position, with the strongest separation at Music Start. The equal-weight mean over the first 128 positions remains highly aligned with positions 32--128 ($0.93$--$0.97$) and with position 256 ($0.84$--$0.86$).

Vocal gender is therefore not represented by one fixed direction throughout the music-token stage: Music Start is distinct, but the representation becomes substantially more stable once generation is underway.

\FloatBarrier

\subsection{Causal Effect of the Vocal-Gender Circuit}
\label{app:vocal-circuit}

The greater stability after Music Start motivates a circuit that aggregates information across early music-token generation. We average the shared-history component differences and residual directions over
positions 0--127 and construct a first-128 circuit with the music-token component
score and budget of \S\ref{sec:klw-steering}: 392 MLP neurons and 168 attention
heads. Because the contrast is binary, one circuit serves both targets with opposite signs, $\delta_F=(F-M)/2$ and $\delta_M=-\delta_F$.

We test the selected components at every position from 0 to 127 on held-out lyrics. Enhancement adds $\delta_F$ or $\delta_M$ under a prefix with no gender descriptor; ablation zeroes the same components under the corresponding gender prefix. Random controls match the number of components in each layer, and the enhancement experiment also includes a random control matched to the target's per-layer injection norm.

\begin{figure}[tbp]
    \centering
    \includegraphics[width=0.94\linewidth]{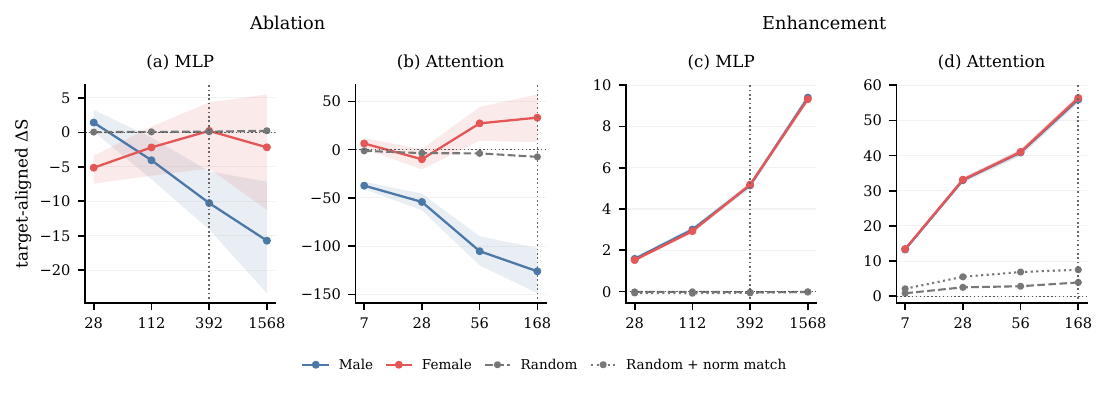}
    \caption{\textbf{Causal intervention on the first-128 vocal-gender circuit.}
    From left to right: MLP ablation, attention-head ablation, MLP enhancement, and attention-head enhancement over positions 0--127.
    The y-axis is target-aligned $\Delta S$ (the male effect is sign-flipped so that positive values always point toward the target).
    Vertical lines mark the generation budgets; bands are 95\% lyric-family bootstrap intervals.
    Dashed gray curves are random controls and dotted gray curves are norm-matched random controls for enhancement.
    Enhancement is specific and nearly symmetric across targets, whereas attention-head ablation is nonspecific.}
    \label{fig:vocal-causal}
\end{figure}

Enhancement gives a clean component-level causal effect (Figure~\ref{fig:vocal-causal}). At the generation budget, the 168 selected heads produce a target-aligned $\Delta S$ of about $56$ for both targets, compared with about $4$ for random heads and $7.5$ for norm-matched random heads. The 392 selected MLP neurons produce a target-aligned shift of about $5.2$, while random neurons remain near zero. All held-out units move in the intended direction.

Ablation is less interpretable. Zeroing the selected attention heads produces large changes toward the female direction under both prefixes, rather than a symmetric loss of the current target, and MLP ablation has the expected sign only under the male prefix. We therefore treat enhancement, rather than ablation, as the clean causal evidence that the selected components can write the vocal-gender direction.

\subsection{Circuit Steering Changes Vocal Gender without Gender Text}
\label{app:vocal-steering}

We finally test whether the circuit changes the generated audio rather than only the internal representation. On the 32 held-out lyrics, all circuit conditions receive the fixed ABC score and
no gender text, acoustic rendering is not intervened on, and the component
difference is injected at every music-token step. The strength of the first-128
variants is selected on four representation-analysis lyrics from
$\{0.5,1.0,1.5\}$ and fixed before evaluation.

We compare the first-128 circuit with the single-position Music-Start circuit.
To separate the effect of averaging the injected difference from the effect of
reselecting components, we also include a hybrid that keeps the Music-Start
components but injects the first-128 mean difference. A matched-random circuit uses the same per-layer component counts and injection norms.

All targeted circuits move the final audio toward the requested gender, while the matched-random intervention does not (Table~\ref{tab:vocal-steering}). The Music-Start circuit gives a target-direction CLAP-margin shift of $+0.024$,
and the hybrid gives $+0.025$, compared with $-0.003$ for matched random
components. The first-128 circuit is also effective ($+0.010$), but is weaker than the Music-Start circuit; the direct gender prompt gives $+0.045$. The same ordering is visible in target hits: 38 for the Music-Start circuit and 34 for the first-128 circuit, versus 31 for matched random components.

Thus, the circuit analysis transfers to vocal gender, but the best intervention is attribute-specific. Unlike the genre score-stage result, averaging the first 128 music-token positions does not improve on the single-position Music-Start circuit, even though the later representations are strongly aligned.

\begin{table}[tbp]
\centering
\small
\setlength{\tabcolsep}{3.6pt}
\setlength{\abovecaptionskip}{0pt}\setlength{\belowcaptionskip}{5pt}
\caption{\textbf{Steering without gender text.}
CLAP target hits on 32 held-out lyrics per target (64 target-labeled evaluations in total) and the paired target-direction shift in the female-minus-male margin relative to the empty-descriptor baseline.
A positive shift means movement toward the requested target.}
\label{tab:vocal-steering}
\begin{tabular}{@{}lrrrc@{}}
\toprule
Condition & Male hits & Female hits & Total & Target-dir. shift \\
\midrule
Empty-descriptor baseline              & 11 & 21 & 32 & $0$ \\
Gender prompt                          & 13 & 28 & 41 & $+0.045$ \\
Music-Start circuit                    & 13 & 25 & 38 & $+0.024$ \\
First-128 circuit                      & 12 & 22 & 34 & $+0.010$ \\
Music-Start components, first-128 difference & 16 & 25 & 41 & $+0.025$ \\
Matched random circuit                 & 10 & 21 & 31 & $-0.003$ \\
\bottomrule
\end{tabular}
\end{table}

\FloatBarrier

\paragraph{Limitations.}
CLAP is an automatic proxy for vocal gender; we do not report human or Gemini judgments. Steering is evaluated on 32 held-out lyrics with one seed, and the selected strength of the first-128 circuits lies at the upper edge of the tested grid. Attention-head ablation is nonspecific and should not be interpreted as a clean inverse intervention. Finally, the direction is defined for the canonical ``male vocal'' and ``female vocal'' wording. Paraphrase experiments show strong wording dependence, so we do not claim a wording-invariant abstract gender axis.

\end{document}